\documentclass[journal]{IEEEtran}
\usepackage{array}
\usepackage{tikz}
\usepackage{pgfplots}
\usepackage{pgfplotstable}
\tikzset{>=latex}

\tikzset{
  currentarrow/.style={
    -{Stealth[length=2.2mm,width=0.8mm]},
    line width=0.6pt,
    line cap=round
  }
}
\pgfplotsset{compat=newest}
\usepgfplotslibrary{patchplots,polar}
\usetikzlibrary{plotmarks}
\usepgfplotslibrary{patchplots}
\usepackage{grffile}
\usetikzlibrary{arrows.meta,calc,positioning,shadings}
\usepackage{algorithm}
\usepackage{algpseudocode}
\usepackage{physics}
\usetikzlibrary{arrows.meta,calc,shapes.geometric}%
\usepackage{standalone}    %
\usepackage{amsmath,amssymb,amsfonts}
\usepackage{graphicx} 

\usepackage{booktabs}
\usepackage{multirow}
\usepackage{siunitx}
\usepackage{adjustbox}
\usepackage{graphicx}
\usepackage[caption=false,font=normalsize,labelfont=sf,textfont=sf]{subfig}
\usepackage{textcomp}
\usepackage{stfloats}
\usepackage{url}
\usepackage{verbatim}
\usepackage[none]{hyphenat}
\usepackage{mathtools}
\usepackage{cite}

\usepackage{xcolor}
\definecolor{panelgreen}{RGB}{170,214,126}

\usepackage{graphicx}

\newcommand{\ts}[1]{\widetilde{#1}}
\newcommand{\tM}[1]{\widetilde{\mathbf{#1}}}

\providecommand*{\M}[1]
{\mathbf#1}        
\providecommand*{\V}[1]{\boldsymbol#1} 
\providecommand*{\UV}[1]{\hat{\boldsymbol#1}}  
\providecommand*{\T}[1]{\mathrm{#1}}  
\providecommand*{\unit}[1]{\ensuremath{\T{\,#1}}}  
\providecommand*{\eu}{\ensuremath{\T{e}}}  
\providecommand*{\ju}{\ensuremath{\T{j}}} 
\providecommand*{\diff}{\operatorname{d}\!}  

\begin{document}

\title{Degrees of Freedom and Beamforming for Large Intelligent Surfaces}

\author{Jiawang Li, Alireza Saberkari, Buon Kiong Lau, Mats Gustafsson

\thanks{Manuscript received \today. This work was supported in part by a Project Grant in ELLIIT 
Call  D, in  part  by  NextG2Com  (grant  no.  2023-00541) 
funded by the VINNOVA program for Advanced Digitalisation, and in part by Swedish Research Council SEE-6GIA 2024-06482.  (Corresponding author: \textit{Jiawang Li}).}%
\thanks{Jiawang Li, Buon Kiong Lau and Mats Gustafsson are with the Department of Electrical and Information Technology, Lund University, 22100 Lund, Sweden (e-mail: {\{jiawang.li, buon\_kiong.lau, mats.gustafsson\}}@eit.lth.se).

Alireza Saberkari is
with the Department of Electrical Engineering, Linköping
University, 58183 Linköping, Sweden. (e-mail: alireza.saberkari@liu.se).
}%
}

\markboth{Degrees of Freedom and Beamforming for Large
Intelligent Surfaces, \today}%
{Li: Degrees of Freedom and Beamforming for Large
Intelligent Surfaces}

\maketitle

\begin{abstract}
Spatial degrees of freedom (DoF), sampling, and beamforming are fundamental to multi-user large intelligent surfaces (LISs), where electromagnetic fields must be shaped, resolved, and focused at multiple near-field locations. This work estimates the number of DoF using closed-form expressions derived from the mutual shadow area for representative LIS configurations. The resulting DoF predictions are validated through numerical singular-value spectra, whose spectral knee points closely match the theoretical estimates.
For line-source configurations, an analytic sampling scheme is developed by partitioning the source or observation line into unit-DoF intervals, enabling the selection of spatial samples. Beamforming results using maximum-ratio transmission and zero-forcing demonstrate that approximately the number of DoF independent beams can be formed. Attempting to exceed this limit results in increased interference and degraded performance.
For surface-based LIS configurations, sampling points are instead determined numerically using the discrete empirical interpolation method. The corresponding beamforming results further confirm that the target region can support approximately as many independent beams as predicted by the DoF analysis. Finally, a polarization-aware study reveals that the electric-field components contribute unequally to the DoF and that the total-field DoF is twice that of a single polarization component.
\end{abstract}

\begin{IEEEkeywords}
Degrees of freedom (DoF), large intelligent surfaces (LISs), mutual shadow area, discrete empirical
interpolation method, polarization-aware study
\end{IEEEkeywords}

\section{Introduction}
\IEEEPARstart{L}{arge} intelligent surfaces (LISs) have emerged as a promising technology for sixth-generation (6G) wireless systems in sub-10 GHz bands~\cite{hu2018beyond1,hu2023design}. Owing to their physically large apertures and controllable electromagnetic responses, LISs can improve spatial selectivity and support near-field beamforming for both communication and wireless power transfer (WPT) applications. In practical LIS deployments, many receivers are located in the near-field region~\cite{selvan2017fraunhofer}, where wavefront curvature and spatial field variations become significant. These features enable spatial focusing, user separation, and energy concentration beyond conventional far-field operation~\cite{bjornson2024towards,shen2020wireless}.

Electromagnetic degrees of freedom (DoF) provide a fundamental measure of the number of independent spatial field modes that can be supported between a source region and an observation region in free space~\cite{franceschetti2017wave,Bucci2025, miller2000communicating,poon2005degrees,decarli2021communication,dardari2020communicating,migliore2006role,jensen2008capacity,puggelli2025maximizing,yuan2024breaking,kuang2025bounds}. 
In LIS systems, this quantity is particularly important because the field over the service region is governed by a finite number of spatial DoF. 
For the considered surface-based LIS application, we focus on line-of-sight (LoS) propagation between the LIS and the observation region, excluding reflected paths. 
The available DoF characterize the spatial control capability of the LIS, determining how many separable field patterns or beams can be synthesized over the service region~\cite{zhang2022beam, li2024multiuser, cui2023nearfieldrainbow, chen2024beamspace}. 
In multi-user scenarios, near-field beamforming can direct energy or information-bearing signals toward different users. 
Sufficient DoF support flexible resource allocation and beam focusing, whereas limited DoF cause spatial coupling among users, beams, or regions. 
Therefore, accurate DoF characterization is essential for evaluating the spatial resolution, channel richness, and field-shaping capability of LIS-enabled systems.

Existing DoF analyses have been developed from several complementary perspectives, including mode counting based on Weyl’s law~\cite{weyl1911asymptotische}, singular-value decompositions of propagation operators~\cite{miller2000communicating,poon2005degrees}, and half-wavelength sampling~\cite{bucci1989degrees,bucci1998representation,maisto2021efficient,maisto2021near}. More recently, geometry-based approaches have shown that the spectral knee of the propagation operator can be predicted analytically from mutual-shadow quantities~\cite{gustafsson2025shadow,gustafsson2025degrees}. However, the connection between such analytical DoF estimates and practical sampling and beamforming design for representative LIS configurations remains unexplored.

In this paper, we study LIS-enabled wireless transmission from the perspective of spatial DoF. The mutual-shadow formulation in~\cite{gustafsson2025shadow} provides a geometric interpretation of the DoF supported between source and observation regions. Here, we apply this formulation to representative LIS deployment scenarios of practical interest, where an LIS mounted on a wall serves users or devices located on a horizontal observation surface. For these configurations, we derive closed-form expressions for the number of DoF and validate the resulting predictions using numerical singular-value spectra.

Based on this DoF characterization, an analytic sampling strategy is first
developed for the 2D line-source configuration, where the selected observation
points are directly obtained from the closed-form DoF expression. These
sampling points are then used for maximum-ratio transmission
(MRT)~\cite{rao2001performance} and zero-forcing
(ZF)~\cite{spencer2004zero} beamforming evaluation, verifying the
beam-related interpretation of DoF in the 2D setting.

The analysis is further extended to a 3D source--observation configuration,
where a discrete empirical interpolation method (DEIM)~\cite{hochman2014reduced} is employed to select
observation points over the planar observation region. The corresponding
beamforming results show that the DoF prediction can also guide spatial
sampling and beam synthesis in the 3D setting. Furthermore, the role of polarization is investigated,
showing its impact on the DoF.

The rest of the paper is organized as follows. Section~II introduces the mutual shadow based DoF characterization for LIS near-field propagation and presents closed-form results for representative 3D surface source configurations. The relationship between sampling points and beamforming is described in Section~III. Section~IV develops closed-form sampling strategies for line sources. Section~V develops DEIM nonuniform point selection sampling for surface sources. Section~VI studies polarization-aware DoF, sampling, and beamforming by decomposing the received electric field into Cartesian components. Finally, Section~VII concludes the paper.

\textit{Notation:} Throughout this paper, boldface letters indicate vectors
and boldface uppercase letters designate matrices. Superscript $(\cdot)^{\T{H}}$ stands for Hermitian transpose.

\section{DoF for LIS Application}\label{sec:dof_lis_application}
An LIS may support communication, WPT, or their joint operation, and therefore needs to shape the electromagnetic field over multiple spatial locations rather than focusing on a
single receiver~\cite{zhang2022beam,li2024multiuser,cui2023nearfieldrainbow,chen2024beamspace}.
In this analysis, an idealized LIS channel is considered, where LoS propagation and perfect channel knowledge are assumed.
Mutual coupling, finite-size element radiation patterns, and receiver sensitivity constraints are not included. Consequently, the obtained results should be interpreted as idealized DoF limits for the considered LIS configurations.

\begin{figure}[t]
  \centering
  \includegraphics[width=0.8\linewidth]{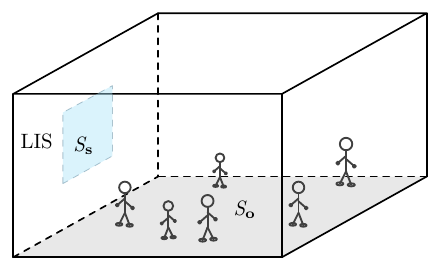}
  \caption{Schematic of the LIS transmitting region $\it S_\T s$ and the users region $\it S_\T o$ for communication or power transfer application. Electric and magnetic current densities are defined on the surface of the LIS $S_\T s$. Only the LoS channel is considered, reflections and inter-user blockage are neglected. $\it S_\T s$ can be a line source or a surface source, corresponding to a linear array or a planar array in a discrete array.}
  \label{fig:lis_receiver}
\end{figure}
We consider a wireless transmission scenario where one or multiple LISs, e.g.,
spatially separated panels in a distributed deployment, are placed on the
transmitting surface $S_\T{s}$ to serve multiple user devices over the receiving
region $S_\T{o}$, as shown in Fig.~\ref{fig:lis_receiver}. 
Although $S_\T{o}$ can in general be chosen as any observation plane of interest parallel to the floor, such as a plane passing through user-held devices, the floor plane is considered here for simplicity.

The available DoF determine the number of independent and distinguishable field patterns or beams that can be controlled by the LIS~\cite{Bucci2025, miller2000communicating, poon2005degrees, decarli2021communication, migliore2006role, jensen2008capacity, puggelli2025maximizing, franceschetti2009mimo}. 
A larger DoF improves spatial selectivity and facilitates the separation of information or energy streams for different users~\cite{decarli2021communication,li2024multiuser,zhang2022beam}, whereas limited DoF may lead to field leakage, spatial coupling, interference, and inefficient resource allocation. 
Under the LoS-only assumption, the obtained DoF should be interpreted as a baseline estimate for the ideal free-space propagation channel, since additional reflected or scattered paths may introduce extra independent spatial modes. 
Therefore, DoF provides a fundamental measure of the capability and limitation of LIS-enabled multi-user communication and WPT systems~\cite{Bucci2025, miller2000communicating, poon2005degrees, decarli2021communication, migliore2006role, jensen2008capacity, puggelli2025maximizing,yuan2024breaking}.

For the continuous case, both the source region $S_\T{s}$ and the observation
region $S_\T{o}$ are treated as continuous spatial domains. The LoS propagation
between them is described by a continuous channel
$\M{H}$. By
performing the singular value decomposition (SVD) of $\M{H}$, we obtain
the singular values of the continuous channel. The $n$th singular value is
denoted by $\sigma_n$. These singular values are arranged in descending order,
where $\sigma_1$ represents the largest singular value.

According to the mutual-shadow-based DoF analysis~\cite{gustafsson2025shadow}, for source region $S_\T s$ and observation area $S_\T o$, the asymptotic number of independent propagation modes is estimated by the total mutual shadow area $\mathcal{A}_{\T{os}}$ between the two regions. In 3D surface settings, this quantity is given by \cite{gustafsson2025shadow}
\begin{equation}
\mathcal{N}_{\T{os}} = \mathcal{A}_{\T{os}} / \lambda^{2}
\label{eq:dof(7)}
\end{equation}
per polarization of the EM field, where $\lambda$ denotes the corresponding wavelength.
In 2D line settings, it is estimated by mutual shadow length $\mathcal{L}_\T{os}$
\begin{equation}
{\mathcal{N}}_{\T{os}} = \mathcal{L}_{\T{os}} / \lambda.
\label{eq:line_source_short}
\end{equation}
These expressions indicate that the number of distinguishable spatial modes scales with the geometrical overlap between the transmitter region and the user region.
 
More specifically, assuming that every point on the transmitting surface is observable from every point on the receiving region, the total mutual shadow area $\mathcal{A}_{\T{os}}$ in 3D can be expressed as \cite{gustafsson2025shadow,brick2026interpreting} (see Fig.~\ref{fig:lis_receiver})
\begin{equation}
\mathcal{A}_{\T{os}}
=
\int_{S_\T{o}}
\int_{S_\T{s}}
\frac{
\left|\hat{\V n}^\prime\cdot (\V r-\V r^\prime)\right|
\left|\hat{\V n} \cdot (\V r-\V r^\prime)\right|
}{
\left|\V r-\V r^\prime\right|^4
}
\, \diff \T S^\prime \diff \T S.
\label{eq:closed_dof_area}
\end{equation}
 and the corresponding 2D mutual shadow length is
\begin{equation}
\mathcal{L}_{\T {os}}
=
\int_{S_\T{o}}
\int_{S_\T{s}}
\frac{
\left|\hat{\V n}^\prime \cdot (\V r-\V r^\prime)\right|
\left|\hat{\V n} \cdot (\V r-\V r^\prime)\right|
}{
|\V r-\V r^\prime|^3
}
\, \diff l^\prime\diff l,
\label{eq:closed_dof_line}
\end{equation}
where $\hat{\V n}'$  and $\hat{\V n}$ denote the unit normal vectors of $S_\T s$ and $S_\T o$, respectively, $\V r^\prime \in S_\T s$ and $\V r \in S_\T o$ denote the position vectors, at the source and observation points, respectively. They are analogous to the view factors used in thermal radiative heat transfer, for which analytical expressions are available for simple configurations~\cite{howell2011radiative}. These expressions explicitly show that the resolvable number of dominant modes depends on the geometry of the transmitter and user regions, their relative separation, and the operating wavelength. Hence, increasing the mutual shadow, or decreasing the wavelength generally increase the number of separable user positions.

\begin{figure}[t]
  \centering \includegraphics[width=\linewidth]{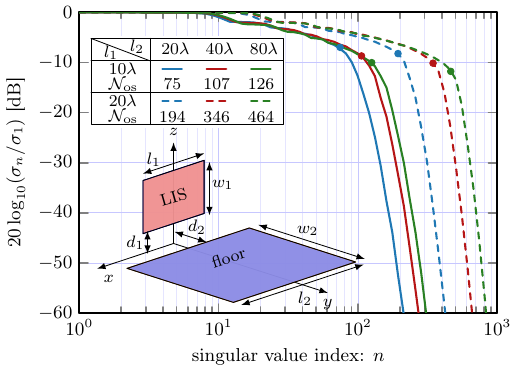}
  \caption{Normalized singular values $\sigma_n/\sigma_1$ between the LIS and the floor on a dB scale, for $l_1=w_1$, $l_2=w_2$, and $d_1=d_2=2\lambda$.
The analytical values obtained from the number of DoF \eqref{eq:dof(7)} and mutual shadow area \eqref{eq:area_LIS_floor} are given in the legend and indicated by markers.
The propagation between the LIS and the floor is evaluated using the scalar
Green's function $G_3$ in App.~\ref{app:2D_Source}. 
}
  \label{fig:numerical_closed_dof}
\end{figure}

To focus on the LIS application, we consider a representative deployment
scenario where an LIS is mounted on a wall to provide wireless communication or
WPT to devices located on the floor in front of it, as illustrated in the inset
of Fig.~\ref{fig:numerical_closed_dof}. For example,
at 6~GHz, side lengths of $10\lambda$ and $20\lambda$ correspond to physical
LIS dimensions of $0.5\unit{m}$ and $1\unit{m}$, respectively, which are
reasonable sizes for practical installation units. The total mutual shadow area in \eqref{eq:closed_dof_area}
provides a closed-form estimate of these DoF, and the numerical singular value
spectra are used to verify the corresponding spectral knee points. The continuous channel operator $\M H$ is represented by a densely sampled channel matrix, where the surfaces and lines used in all examples are discretized with five points per wavelength~\cite{gustafsson2025shadow,miller2000communicating}. This equidistantly $\lambda/5$ sampling interval is sufficiently fine and is therefore used as the continuous-channel reference in the following analysis.
For simplicity, we first use the common scalar propagation model~\cite{Bucci2025} based on \(G_3\) in App.~\ref{app:2D_Source}, as an initial example. Polarization effects are considered later in Section~\ref{sec:Polarization-Aware Beamforming}.

Fig.~\ref{fig:numerical_closed_dof} shows the normalized singular value spectra of the propagation operator between the LIS and the floor region for different geometric configurations. A knee behavior can be observed in all curves: the normalized singular values remain significant for the dominant modes and then decay rapidly after a certain index. This knee point indicates the number of propagation modes, i.e., the spatial DoF supported by the corresponding geometry. Moreover, as the dimensions of the LIS and floor regions increase, the knee shifts to the right, which means that more independent spatial modes can be supported. This observation is consistent with the mutual shadow area interpretation \eqref{eq:closed_dof_area}, indicating that the number of DoF is closely related to the geometric coupling between the two regions.

For the geometry shown in Fig.~\ref{fig:numerical_closed_dof}, the mutual shadow area \eqref{eq:closed_dof_area} is evaluated from the view factors in~\cite{howell2011radiative} as
\begin{multline}
\mathcal{A}_\T{os}
=
\Psi(d_2,d_1)
-
\Psi(d_2,d_1+w_1)
-
\Psi(d_2+w_2,d_1) \\
+
\Psi(d_2+w_2,d_1+w_1),
\label{eq:area_LIS_floor}
\end{multline}
where $\Psi(\alpha,\beta)=\Phi(\sqrt{\alpha^2+\beta^2},s_-)-\Phi(\sqrt{\alpha^2+\beta^2},s_+)$,
\begin{equation}
\Phi(\gamma,s)
=
s\gamma\,
\arctan\left(
\frac{s}{\gamma}
\right)
-
\frac{\gamma^2-s^2}{4}
\ln\left(\gamma^2+s^2\right),
\label{eq:area_LIS_floor_2}
\end{equation}
and
$s_\pm={|l_2\pm l_1|}/{2}.$ The four terms in \eqref{eq:area_LIS_floor} account for the contributions from the four corner combinations of the two rectangular regions, and their alternating signs follow from the inclusion--exclusion form of the view-factor expression~\cite{howell2011radiative}. The corresponding number of spatial DoF is then obtained as $\mathcal{N}_\T{os}$~\eqref{eq:dof(7)}, which is used as the reference value for the sampling and beamforming results for the examples in Section~\ref{sec:3D Surface Source} and Section~\ref{sec:Polarization-Aware Beamforming}.

The solid markers in Fig.~\ref{fig:numerical_closed_dof} indicate the analytical DoF values \eqref{eq:dof(7)} estimated by the mutual shadow area \eqref{eq:area_LIS_floor} and \eqref{eq:area_LIS_floor_2} by taking
the ceiling of the DoF. 
These values approximately locate the knee points of the numerical singular value spectra, showing that the closed-form expression can estimate the number of DoF\cite{gustafsson2025shadow,brick2026interpreting}.

\section{Beam-Aware Receiving Sampling Distribution}\label{section 3}
Building on the DoF analysis in Section~\ref{sec:dof_lis_application}, 
receiver-plane sampling is introduced to convert the continuous observation
region into a finite set of representative target points. 
This allows the estimated spatial DoF to be directly related to practical
beamforming performance, such as how many independent beams can be formed. The sampling analysis evaluates how many spatially separable beams can be supported within the observation region, considering focal-spot locations, peak received field intensity, main-lobe width, and sidelobe leakage.

To make the link between the DoF analysis and practical beamforming evaluation explicit, we distinguish between the densely sampled channel operator $\M H$ used for DoF analysis and the sampling channel matrix $\tM H$ used for beamforming analysis. Specifically, $\M H$ is constructed on a dense grid over the receiver observation region, while $\tM H$ is obtained by retaining only the rows of $\M H$ corresponding to the selected receiver-plane sampling points.

The singular value spectrum of $\tM{H}$ provides a discrete
representation of the spatial modes that can be excited and observed at the
selected receiver-plane samples. The singular
values $\ts{\sigma}_n$ is obtained from the SVD
of the sampled channel matrix
$
\tM{H}
=
\M U\M \Sigma \M V^{\T H}$. The diagonal matrix
$\boldsymbol{\Sigma}=\operatorname{diag}(\ts{\sigma}_1,\ts{\sigma}_2,\ldots,\ts{\sigma}_{N_\T o})$
contains the singular values.

According to the singular-value interlacing inequalities \cite{horn2012matrix},
if $\tM{H}$ is a row submatrix of $\M H$ when
$p$ rows are removed from $\M H$ to form
$\widetilde{\M H}$, the singular values satisfy the interlacing relation\cite{horn2012matrix}
\begin{equation}
\sigma_i(\M H)
\geq
\ts{\sigma}_i(\tM{H})
\geq
\sigma_{i+p}(\M H),
\label{eq:sv_interlacing}
\end{equation}
This inequality shows that
receiver plane sampling cannot create additional independent spatial modes
beyond those supported by the full channel. Instead, the purpose of the
sampling strategy is to preserve the dominant singular modes associated with
the spatial DoF. Consequently, the knee of the singular-value
spectrum of $\tM{H}$ provides an indicator of how many separable beams can be reliably supported.

Based on the sampled channel matrix, representative beamforming strategies
are then used to examine how the available spatial DoF can be converted into
practical focal beams. In particular, MRT \cite{rao2001performance} and
ZF \cite{spencer2004zero}  beamforming are
considered.

The MRT beamformer~\cite{rao2001performance} uses the conjugated (time-reversed) channel response as the
beamforming weight matrix
\begin{equation}
\M W_{\T{MRT}}
=
\tM{H}^{\T H}=\M V\M \Sigma \M U^{\T H}.
\label{eq:mrt_weight}
\end{equation}
The MRT mainly evaluates the ability of the aperture to enhance the field at a desired sampling point.

The ZF beamformer aims to synthesize independent beams at the selected
sampling points. Ideally, it satisfies
$
\tM{H}\M W_{\rm ZF}
=
\M I_{N_{\T o}}.
$ When $N_{\T s}\geq N_{\T o}$ and $\tM{H}$ has full row rank,
the minimum-norm ZF beamforming weight matrix is
\begin{equation}
\M W_{\rm ZF}
=
\tM{H}^{\T H}
\left(
\tM{H}\tM{H}^{\T H}
\right)^{-1} = \M V\M \Sigma^{-1} \M U^{\T H}.
\label{eq:Wzf}
\end{equation}
 However, because the strict ZF solution in \eqref{eq:Wzf} involves the inversion of $\M\Sigma$, small singular values beyond the spectral knee can be strongly amplified and therefore reduce robustness. This issue can be alleviated, for example, by minimum mean-square-error (MMSE) regularization \cite{christensen2008weighted}, which reduces the influence of weakly excited modes beyond the knee of the singular-value spectrum.

The performance of MRT~\eqref{eq:mrt_weight} and ZF~\eqref{eq:Wzf} beamforming is closely tied to the singular-value spectrum of $\tM{H}$. In MRT, the beam response is naturally dominated by the strongest singular modes, whereas modes beyond the spectral knee contribute progressively less field energy. Consequently, the achievable dynamic range after retaining $n$ modes is governed by $\sigma_n/\sigma_1$. For ZF beamforming, the impact of the singular-value distribution is even more pronounced. Since the inverse operation scales the $n$th singular mode by $1/\sigma_n$, small singular values result in large beamforming weights, increased transmit-power requirements, and greater sensitivity to noise and model imperfections. In this case, the relevant dynamic range becomes $\sigma_1/\sigma_n$. As a result, both MRT and ZF exhibit the same knee observed in the singular-value spectrum. This knee marks the transition from dominant propagating modes to weakly excited modes and provides a beamforming-based interpretation of the DoF of the channel.

To ensure a fair comparison under equal excitation level, each beamforming
vector~\eqref{eq:mrt_weight},~\eqref{eq:Wzf} is normalized as
\begin{equation}
\overline{\V w}_n
={\V w_n}/{\|\V w_n\|_2},
\qquad
n=1,\ldots,N_{\T o},
\label{eq:column_power_normalization}
\end{equation}
The beams shown in
the subsequent sections are all normalized according to this method.

\section{2D Line Source}\label{SEC:2D}
\begin{figure}[t]
  \centering  \includegraphics[width=\linewidth]{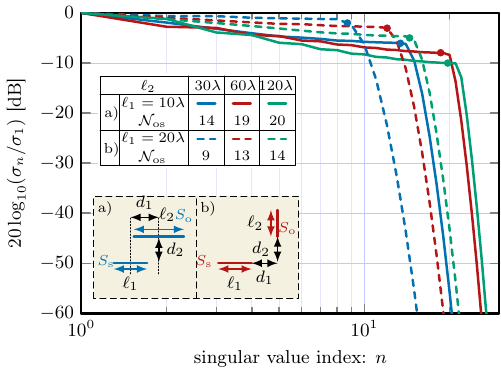}
  \caption{Normalized singular values $\sigma_n/\sigma_1$ for two 2D line source configurations on a dB scale under $d_1=d_2=10\lambda$. The analytical values
obtained from number of DoF \eqref{eq:line_source_short} and mutual shadow length \eqref{eq:closed_dof_line} are given in the legend and indicated by markers. The channel matrix is evaluated using Green’s function $G_2$ in \eqref{eq:transmission matrices}.}  \label{fig:2D_line_source}
\end{figure}

Two representative sets of transmitting and receiving line sources with specific relative arrangements are considered. In the first case, the two line sources are parallel. In the second case, the two line sources are orthogonal to each other without physical contact. Their relative arrangement and parametric representations are illustrated in Fig.~\ref{fig:2D_line_source}. The used 2D Green's function $G_2$ is provided in App.~\ref{app:2D_Source}.

\subsection{DoF Analysis}
The mutual shadow length $\mathcal{L}_{\T{os}}$ for parallel transmitter $S_\T s$ and receiver $S_\T o$ lines
 (see Fig.~\ref{fig:2D_line_source}a), is
\begin{equation}
\mathcal{L}_{\T{os}}^{\parallel}=f(\delta_+)-f(\delta_-).
\label{eq:line_source_offset}
\end{equation}
where $\delta_\pm=\lvert \ell_1\pm\ell_2\rvert$ and
\begin{equation}
f(\delta_\pm)=\sqrt{d_2^2+(\delta_\pm/2-d_1)^2}
+\sqrt{d_2^2+(\delta_\pm/2+d_1)^2}.
\label{eq:f(x)}
\end{equation}

Another interesting geometry is the placement of the line source, similar to that in Fig.~\ref{fig:lis_receiver}. For the configuration given in Fig.~\ref{fig:2D_line_source}b, the corresponding mutual shadow length is given by
\begin{equation}
\mathcal{L}_{\T{os}}^{\perp}=
F(z_1)
- F(z_2).
\label{eq:LTR_perp}
\end{equation}
where 
\begin{equation}
\begin{multlined}
F(z)=\sqrt{z^2+\left({\ell_1}+d_1\right)^2}-\sqrt{z^2+d_1^2}
.
\end{multlined}
\label{eq:F1F2}
\end{equation}
$z_1=d_2$, and $z_2=d_2+\ell_2$.

For Fig.~\ref{fig:2D_line_source}, enlarging the receiving line source shifts the knee of the singular-value spectrum to a larger index, but the increase in spatial modes is not proportional to $\ell_2$. When $\ell_2$ increases from $60\lambda$ to $120\lambda$, the knee position changes only slightly, showing a gradual saturation of the DoF. This saturation behavior can also be observed directly from the closed-form mutual-shadow expressions. In the parallel configuration, when the receiving line becomes infinitely long, i.e., $\ell_2\to\infty$, the offset parameters become $\delta_+=\ell_2+\ell_1$ and $\delta_-=\ell_2-\ell_1$. The mutual shadow length converges to
\begin{equation}
\lim_{\ell_2\to\infty}
\mathcal{L}_{\T{os}}^{\parallel}=
\lim_{\ell_2\to\infty}
\left[
f(\ell_2+\ell_1)-f(\ell_2-\ell_1)
\right]=
2\ell_1 .
\label{eq:parallel_infinite_rx_limit}
\end{equation}
Noticed that we get $\mathcal{N}_\T{os}=\ell_1/(\lambda/2)$ when $\ell_2\to\infty$, which is consistent with the Nyquist–Shannon  $\lambda/2$
sampling rule~\cite{franceschetti2017wave}. For the perpendicular configuration, the second term $F(d_2+\ell_2)$ vanishes as $\ell_2\to\infty$. Therefore,
\begin{equation}
\lim_{\ell_2\to\infty}
\mathcal{L}_{\T{os}}^{\perp}=
\sqrt{d_2^2+\left(\ell_1+d_1\right)^2}-
\sqrt{d_1^2+d_2^2}.
\label{eq:perp_infinite_rx_limit}
\end{equation}
These two limiting results show that increasing the receiving-line length cannot indefinitely increase the number of effective spatial modes. The parallel geometry is ultimately limited by the effective transmitting-line length, whereas the perpendicular geometry is further constrained by the relative distances $d_1$ and $d_2$.

\subsection{Sampling}
This subsection explains how the DoF estimate is converted into beamforming
sampling points. The basic idea is to use the theoretical DoF as the number of
resolvable focal positions on the receiving line. Instead of sampling the
receiving line only according to its physical length, the sampling points are
selected according to the mutual-shadow length. In this
way, the continuous receiving line is represented by a finite set of focal
positions determined by the available spatial DoF.

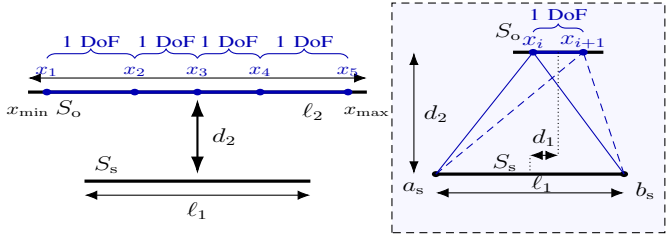
\begin{figure}[t]
    \centering
    \begin{tikzpicture}[
    xscale=0.83,
    yscale=0.62,
    transform shape,
    >=Latex,
    font=\large,
    thick,
    sample/.style={circle,fill=black,inner sep=1.5pt},
    xsample/.style={circle,fill=blue!70!black,inner sep=1.4pt}
]

\def\xL{0.8}
\def\xR{6.2}
\def\xC{3.5}

\def\xA{1.10}
\def\xB{2.50}
\def\xCc{3.50}
\def\xD{4.50}
\def\xE{5.90}


\draw[very thick] (\xL,3.0) -- (\xR,3.0);
\node[above=4pt] at (1.45,2.2) {$S_\mathrm{o}$};

\draw[very thick] (1.7,1.1) -- (5.3,1.1);
\node[below=4pt] at (2.05,1.9) {$S_\mathrm{s}$};

\draw[<->,thin] (\xL,3.30) --
node[pos=0.88,above=1pt,xshift=-6pt,yshift=-30pt] {$\ell_2$}
(\xR,3.30);

\draw[<->,thin] (1.7,0.80) -- node[below=1pt] {$\ell_1$} (5.3,0.80);

\draw[<->] (\xC,1.25) -- node[right=3pt] {$d_2$} (\xC,2.85);

\node[below=5pt,font=\normalsize] at (\xL,3.0) {$x_{\min}$};
\node[below=5pt,font=\normalsize] at (\xR,3.0) {$x_{\max}$};


\foreach \x in {\xA,\xB,\xCc,\xD,\xE}{
    \fill[blue!70!black] (\x,3.0) circle (2pt);
}

\node[above=6pt,blue!70!black,font=\normalsize] at (\xA,3.0) {$x_1$};
\node[above=6pt,blue!70!black,font=\normalsize] at (\xB,3.0) {$x_2$};
\node[above=6pt,blue!70!black,font=\normalsize] at (\xCc,3.0) {$x_3$};
\node[above=6pt,blue!70!black,font=\normalsize] at (\xD,3.0) {$x_4$};
\node[above=6pt,blue!70!black,font=\normalsize] at (\xE,3.0) {$x_5$};


\draw[blue!70!black,line width=1.2pt] (\xA,3.0) -- (\xB,3.0);
\draw[blue!70!black,line width=1.2pt] (\xB,3.0) -- (\xCc,3.0);
\draw[blue!70!black,line width=1.2pt] (\xCc,3.0) -- (\xD,3.0);
\draw[blue!70!black,line width=1.2pt] (\xD,3.0) -- (\xE,3.0);


\draw[
    decorate,
    decoration={brace,amplitude=3.5pt},
    blue!70!black,
    thin
]
(\xA,3.72) -- node[above=3pt,font=\normalsize,blue!70!black] {1 DoF} (\xB,3.72);

\draw[
    decorate,
    decoration={brace,amplitude=3.5pt},
    blue!70!black,
    thin
]
(\xB,3.72) -- node[above=3pt,font=\normalsize,blue!70!black] {1 DoF} (\xCc,3.72);

\draw[
    decorate,
    decoration={brace,amplitude=3.5pt},
    blue!70!black,
    thin
]
(\xCc,3.72) -- node[above=3pt,font=\normalsize,blue!70!black] {1 DoF} (\xD,3.72);

\draw[
    decorate,
    decoration={brace,amplitude=3.5pt},
    blue!70!black,
    thin
]
(\xD,3.72) -- node[above=3pt,font=\normalsize,blue!70!black] {1 DoF} (\xE,3.72);

\begin{scope}[xshift=6.6cm]

\fill[blue!3] (0,0.0) rectangle (4.35,4.9);
\draw[densely dashed,thin] (0,0.0) rectangle (4.35,4.9);

\draw[very thick] (0.7,1.25) -- (3.7,1.25);
\node[below=4pt] at (1.8,1.95) {$S_\mathrm{s}$};

\fill (0.7,1.25) circle (1.8pt);
\fill (3.7,1.25) circle (1.8pt);
\node[below left=2pt] at (0.7,1.25) {$a_\mathrm{s}$};
\node[below right=2pt] at (3.7,1.25) {$b_\mathrm{s}$};

\draw[<->,thin] (0.7,0.85) -- node[above right=-4pt] {$\ell_1$} (3.7,0.85);

\def\xri{2.25}
\def\xrj{3.05}
\def\yr{3.85}

\pgfmathsetmacro{\xrc}{(\xri+\xrj)/2}

\draw[very thick] ({\xri-0.32},\yr) -- ({\xrj+0.32},\yr);
\node[above=4pt] at (\xrc-0.8,\yr) {$S_\mathrm{o}$};

\draw[blue!70!black,line width=1.2pt] (\xri,\yr) -- (\xrj,\yr);

\fill[blue!70!black] (\xri,\yr) circle (2pt);
\fill[blue!70!black] (\xrj,\yr) circle (2pt);

\node[above=4pt,blue!70!black] at (\xri,\yr-0.18) {$x_i$};
\node[above=4pt,blue!70!black] at (\xrj,\yr-0.18) {$x_{i+1}$};

\draw[
    decorate,
    decoration={brace,amplitude=3pt},
    blue!70!black,
    thin
]
(\xri,4.28) -- node[above=4pt,font=\normalsize,blue!70!black] {1 DoF} (\xrj,4.28);


\draw[<->,thin] (0.35,1.25) -- node[right=1pt] {$d_2$} (0.35,\yr);

\draw[densely dotted,thin] (2.2,1.25) -- (2.2,1.65);
\draw[densely dotted,thin] (\xrc,\yr) -- (\xrc,1.65);
\draw[<->,thin] (2.2,1.65) -- node[above=2pt] {$d_1$} (\xrc,1.65);

\draw[blue!70!black,thin] (0.7,1.25) -- (\xri,\yr);
\draw[blue!70!black,thin] (3.7,1.25) -- (\xri,\yr);

\draw[blue!70!black,thin,densely dashed] (0.7,1.25) -- (\xrj,\yr);
\draw[blue!70!black,thin,densely dashed] (3.7,1.25) -- (\xrj,\yr);

\end{scope}

\end{tikzpicture}
    \caption{Illustration of the sampling process for   Fig.~\ref{fig:2D_line_source}a. The receiving line is divided according to
    the available DoF, and the representative focal positions are selected from
    the midpoint of each resolvable interval. The right subfigure provides a close-up view of one resolvable interval.}
    \label{fig:sampling}
\end{figure}

As illustrated in Fig.~\ref{fig:sampling}, the receiving line is divided into
several resolvable intervals according to the DoF estimate, and one
representative focal position is selected from the midpoint of each interval.
The interval boundaries are determined such that the DoF contribution of each
interval is one, i.e., \(\mathcal N_{\T{os}}=1\) for each interval. Therefore, each
interval corresponds to one independent spatial mode that can be resolved
between the transmitting and receiving lines. Therefore, the DoF estimate determines the number of representative focal
positions used for beamforming on the receiving line.

In the parallel case, after
\(\mathcal{L}_{\T{os}}^{\parallel}\) is evaluated from
\eqref{eq:line_source_offset}, this effective interval is divided into
\(\mathcal{N}_{\T{os}}^{\parallel}\) resolvable parts. The midpoint of each
part is then converted to a physical position on the receiving line through the
same endpoint-distance relation used in \(f(\cdot)\).

In the perpendicular case, the procedure is applied to the interval
\(F(z_1)-F(z_2)\). This interval is divided into
\(\mathcal{N}_{\T{os}}^{\perp}\) resolvable parts, and each midpoint is mapped
back to the corresponding physical depth by solving the inverse relation of
\(F(z)\) over \(z\in[z_1,z_2]\). The resulting points form the beamforming
focal positions along the receiving line.

\subsection{Beamforming}

\begin{figure}[t]
  \centering
\includegraphics[width=\linewidth]{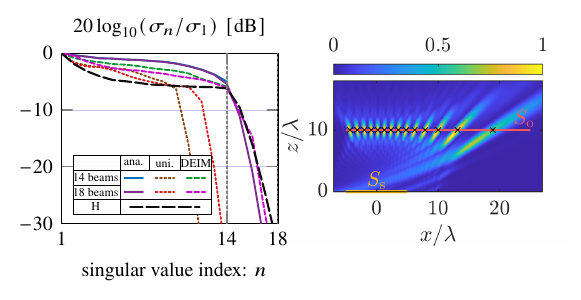}
  \caption{Normalized singular values and beamforming intensity distribution for the line-source configuration in Fig.~\ref{fig:2D_line_source}a. 
The left panel compares 14, 18 beams at sampling point positions using analytical, uniform sampling and the DEIM method. \(\M H\) denotes the result obtained from the continuous source and observation regions.
The right panel shows the corresponding 2D normalized field distribution using MRT method~\eqref{eq:mrt_weight}, where the black markers denote the prescribed beam locations.}  \label{fig:2D_line_svd_field}
\end{figure}

To evaluate the beamforming performance obtained from the DoF-guided sampling
scheme, we consider the configuration shown in
Fig.~\ref{fig:2D_line_source}a. For the beam verification, we use
\(\ell_2=3\ell_1=3d_1=30\lambda\).

We first examine the singular-value behavior of the sampled channel, as shown
in the left panel of Fig.~\ref{fig:2D_line_svd_field}. The $\mathcal N_\T{os}\approx 14$-beam analytical
sampling case is consistent with the DoF estimates in
\eqref{eq:line_source_short} and \eqref{eq:line_source_offset}, and its
spectral transition approximately agrees with the knee position of the
reference channel \(\M H\).

The effect of prescribing more beams than the estimated DoF is also shown in
the same panel. Compared with the estimated DoF of 14, the 18-beam cases
introduce additional sampling constraints beyond the reliable spatial modes.
This leads to a sharper spectral decay after the knee and indicates that the
extra beams are not supported as independent controllable modes. 

To select representative sampling locations numerically, we also employ the
DEIM method~\cite{hochman2014reduced}. A dense candidate channel matrix is
first constructed over the observation region, and its dominant left singular
vectors are extracted by SVD. The retained DEIM rank is determined by estimated DoF \(\mathcal N_{\T{os}}\approx 14\) in
\eqref{eq:line_source_offset}. The DEIM greedy method is then applied to the
retained singular-vector subspace, and the selected row indices are mapped to
the corresponding physical sampling locations on \(S_{\mathrm{o}}\). This
directly relates the number of sampling points to the spatial DoF,
while DEIM identifies representative locations from the dominant singular-vector
subspace instead of relying on uniformly distributed samples.

In the uniformly distributed sampling case, the sampling points are uniformly distributed between the two endpoints of the observation region, including both endpoints. For the same
number of 18 beams, the DEIM-based sampling gives a more stable singular-value
distribution before the spectral knee than uniform sampling, suggesting that
DEIM selects more representative sampling locations from the dominant modal
subspace. In contrast, the singular-value spectrum obtained with uniform sampling shows
an earlier knee, indicating that the uniform sampling distribution cannot
preserve as many dominant modes as the continuous spectrum. This confirms that the DoF limits the number of distinguishable
beams that can be robustly synthesized.
It is worth noting that, among the three schemes, the proposed analytical method~\eqref{eq:line_source_offset} preserves the same knee position in the singular-value spectrum as the reference channel matrix $\M{H}$. Meanwhile, the channel singular values before the knee are more uniformly distributed, indicating the superiority of the proposed method.

The right panel of Fig.~\ref{fig:2D_line_svd_field} shows the corresponding
2D normalized field distribution for 14 beams. The high-intensity regions
are concentrated around the selected focal positions, indicating that the
DoF-guided sampling points are physically meaningful and correspond to
controllable focusing locations in the observation region.

\begin{figure}[t]
  \centering
\definecolor{c1}{named}{blue}
\definecolor{c2}{named}{red}
\definecolor{c3}{rgb}{0,0.5,0.5}
\definecolor{c4}{rgb}{1,0.45,0}
\definecolor{c5}{named}{violet}
\definecolor{c6}{rgb}{0,0.45,0}
\definecolor{c7}{named}{magenta}
\definecolor{c8}{rgb}{0,0.55,0.65}
\definecolor{c9}{rgb}{0.45,0.22,0.05}
\definecolor{c10}{rgb}{0.9,0.25,0.55}
\definecolor{c11}{rgb}{0.45,0.8,0}
\definecolor{c12}{rgb}{0.45,0.45,0}
\definecolor{c13}{rgb}{0.45,0.15,0.65}
\definecolor{c14}{rgb}{0.45,0.45,0.45}
\definecolor{c15}{named}{black}
\definecolor{c16}{rgb}{0,0.65,0.9}
\definecolor{c17}{rgb}{0.9,0.25,0}
\definecolor{c18}{rgb}{0.25,0.25,0.75}

\definecolor{numOdd}{named}{blue}
\definecolor{numEven}{named}{red}

\begin{tikzpicture}[
    line join=round,
    line cap=round,
    font=\fontsize{9}{11}\selectfont
]

\begin{axis}[
    name=upperplot,
    at={(0cm,2.20cm)},
    anchor=south west,
    width=7.08cm,
    height=2cm,
    scale only axis,
    xmin=-5.5,
    xmax=25.5,
    ymin=0,
    ymax=1.12,
    xtick={-5,0,5,10,15,20,25},
    xticklabels={},
    ytick={0,0.2,0.4,0.6,0.8,1.0},
    ylabel={norm. amp.},
    xlabel={},
    ylabel style={font=\fontsize{9}{11}\selectfont},
    xticklabel style={font=\fontsize{9}{11}\selectfont},
    yticklabel style={font=\fontsize{9}{11}\selectfont},
    axis background/.style={fill=white},
    grid=major,
    major grid style={thin,blue!15!white},
    tick align=inside,
    line width=1pt,
    tick style={line width=0.8pt},
    clip=false
]

\addplot[c1,  line width=0.8pt] table[x index=0, y index=1]  {data/beam_width_ZF.dat};
\addplot[c2,  line width=0.8pt] table[x index=0, y index=2]  {data/beam_width_ZF.dat};
\addplot[c3,  line width=0.8pt] table[x index=0, y index=3]  {data/beam_width_ZF.dat};
\addplot[c4,  line width=0.8pt] table[x index=0, y index=4]  {data/beam_width_ZF.dat};
\addplot[c5,  line width=0.8pt] table[x index=0, y index=5]  {data/beam_width_ZF.dat};
\addplot[c6,  line width=0.8pt] table[x index=0, y index=6]  {data/beam_width_ZF.dat};
\addplot[c7,  line width=0.8pt] table[x index=0, y index=7]  {data/beam_width_ZF.dat};
\addplot[c8,  line width=0.8pt] table[x index=0, y index=8]  {data/beam_width_ZF.dat};
\addplot[c9,  line width=0.8pt] table[x index=0, y index=9]  {data/beam_width_ZF.dat};
\addplot[c10, line width=0.8pt] table[x index=0, y index=10] {data/beam_width_ZF.dat};
\addplot[c11, line width=0.8pt] table[x index=0, y index=11] {data/beam_width_ZF.dat};
\addplot[c12, line width=0.8pt] table[x index=0, y index=12] {data/beam_width_ZF.dat};
\addplot[c13, line width=0.8pt] table[x index=0, y index=13] {data/beam_width_ZF.dat};
\addplot[c14, line width=0.8pt] table[x index=0, y index=14] {data/beam_width_ZF.dat};

\node[anchor=center, font=\fontsize{6.5}{7}\selectfont, text=black] at (axis cs:-4.342356286919999,1.03) {1};
\node[anchor=center, font=\fontsize{6.5}{7}\selectfont, text=black] at (axis cs:-3.119510095320000,1.03) {2};
\node[anchor=center, font=\fontsize{6.5}{7}\selectfont, text=black] at (axis cs:-1.979415354400000,1.03) {3};
\node[anchor=center, font=\fontsize{6.5}{7}\selectfont, text=black] at (axis cs:-0.886676660540000,1.03) {4};
\node[anchor=center, font=\fontsize{6.5}{7}\selectfont, text=black] at (axis cs:0.186025300580000,1.03) {5};
\node[anchor=center, font=\fontsize{6.5}{7}\selectfont, text=black] at (axis cs:1.262873195260000,1.03) {6};
\node[anchor=center, font=\fontsize{6.5}{7}\selectfont, text=black] at (axis cs:2.368685231880000,1.03) {7};
\node[anchor=center, font=\fontsize{6.5}{7}\selectfont, text=black] at (axis cs:3.532916472260000,1.03) {8};
\node[anchor=center, font=\fontsize{6.5}{7}\selectfont, text=black] at (axis cs:4.795444418360000,1.03) {9};
\node[anchor=center, font=\fontsize{6.5}{7}\selectfont, text=black] at (axis cs:6.217049693320000,1.03) {10};
\node[anchor=center, font=\fontsize{6.5}{7}\selectfont, text=black] at (axis cs:7.901915102559999,1.03) {11};
\node[anchor=center, font=\fontsize{6.5}{7}\selectfont, text=black] at (axis cs:10.054957102219998,1.03) {12};
\node[anchor=center, font=\fontsize{6.5}{7}\selectfont, text=black] at (axis cs:13.166002301019999,1.03) {13};
\node[anchor=center, font=\fontsize{6.5}{7}\selectfont, text=black] at (axis cs:18.891239312019998,1.03) {14};

\end{axis}

\begin{axis}[
    name=lowerplot,
    at={(0cm,0cm)},
    anchor=south west,
    width=7.08cm,
    height=2cm,
    scale only axis,
    xmin=-5.5,
    xmax=25.5,
    ymin=0,
    ymax=1.12,
    xtick={-5,0,5,10,15,20,25},
    ytick={0,0.2,0.4,0.6,0.8,1.0},
    ylabel={norm. amp.},
    xlabel={$x/\lambda$},
    xlabel style={font=\fontsize{9}{11}\selectfont},
    ylabel style={font=\fontsize{9}{11}\selectfont},
    xticklabel style={font=\fontsize{9}{11}\selectfont},
    yticklabel style={font=\fontsize{9}{11}\selectfont},
    axis background/.style={fill=white},
    grid=major,
    major grid style={thin,blue!15!white},
    tick align=inside,
    line width=1pt,
    tick style={line width=0.8pt},
    clip=false
]

\addplot[c1,  dash pattern=on 1.2pt off 1.0pt, line width=0.8pt] table[x index=0, y index=1]  {data/beam_width_ZF_DEIM.dat};
\addplot[c2,  dash pattern=on 1.2pt off 1.0pt, line width=0.8pt] table[x index=0, y index=2]  {data/beam_width_ZF_DEIM.dat};
\addplot[c3,  dash pattern=on 1.2pt off 1.0pt, line width=0.8pt] table[x index=0, y index=3]  {data/beam_width_ZF_DEIM.dat};
\addplot[c4,  dash pattern=on 1.2pt off 1.0pt, line width=0.8pt] table[x index=0, y index=4]  {data/beam_width_ZF_DEIM.dat};
\addplot[c5,  dash pattern=on 1.2pt off 1.0pt, line width=0.8pt] table[x index=0, y index=5]  {data/beam_width_ZF_DEIM.dat};
\addplot[c6,  dash pattern=on 1.2pt off 1.0pt, line width=0.8pt] table[x index=0, y index=6]  {data/beam_width_ZF_DEIM.dat};
\addplot[c7,  dash pattern=on 1.2pt off 1.0pt, line width=0.8pt] table[x index=0, y index=7]  {data/beam_width_ZF_DEIM.dat};
\addplot[c8,  dash pattern=on 1.2pt off 1.0pt, line width=0.8pt] table[x index=0, y index=8]  {data/beam_width_ZF_DEIM.dat};
\addplot[c9,  dash pattern=on 1.2pt off 1.0pt, line width=0.8pt] table[x index=0, y index=9]  {data/beam_width_ZF_DEIM.dat};
\addplot[c10, dash pattern=on 1.2pt off 1.0pt, line width=0.8pt] table[x index=0, y index=10] {data/beam_width_ZF_DEIM.dat};
\addplot[c11, dash pattern=on 1.2pt off 1.0pt, line width=0.8pt] table[x index=0, y index=11] {data/beam_width_ZF_DEIM.dat};
\addplot[c12, dash pattern=on 1.2pt off 1.0pt, line width=0.8pt] table[x index=0, y index=12] {data/beam_width_ZF_DEIM.dat};
\addplot[c13, dash pattern=on 1.2pt off 1.0pt, line width=0.8pt] table[x index=0, y index=13] {data/beam_width_ZF_DEIM.dat};
\addplot[c14, dash pattern=on 1.2pt off 1.0pt, line width=0.8pt] table[x index=0, y index=14] {data/beam_width_ZF_DEIM.dat};
\addplot[c15, dash pattern=on 1.2pt off 1.0pt, line width=0.8pt] table[x index=0, y index=15] {data/beam_width_ZF_DEIM.dat};
\addplot[c16, dash pattern=on 1.2pt off 1.0pt, line width=0.8pt] table[x index=0, y index=16] {data/beam_width_ZF_DEIM.dat};
\addplot[c17, dash pattern=on 1.2pt off 1.0pt, line width=0.8pt] table[x index=0, y index=17] {data/beam_width_ZF_DEIM.dat};
\addplot[c18, dash pattern=on 1.2pt off 1.0pt, line width=0.8pt] table[x index=0, y index=18] {data/beam_width_ZF_DEIM.dat};

\node[anchor=center, font=\fontsize{6.3}{7}\selectfont, text=numOdd]  at (axis cs:-4.992500000000000,1.03) {1};
\node[anchor=center, font=\fontsize{6.3}{7}\selectfont, text=numEven] at (axis cs:-4.782500000000000,1.03) {2};
\node[anchor=center, font=\fontsize{6.3}{7}\selectfont, text=numOdd]  at (axis cs:-3.927500000000000,1.03) {3};
\node[anchor=center, font=\fontsize{6.3}{7}\selectfont, text=numEven] at (axis cs:-3.087500000000000,1.03) {4};
\node[anchor=center, font=\fontsize{6.3}{7}\selectfont, text=numOdd]  at (axis cs:-2.322500000000000,1.03) {5};
\node[anchor=center, font=\fontsize{6.3}{7}\selectfont, text=numEven] at (axis cs:-1.242500000000000,1.03) {6};
\node[anchor=center, font=\fontsize{6.3}{7}\selectfont, text=numOdd]  at (axis cs:0.032500000000001,1.03) {7};
\node[anchor=center, font=\fontsize{6.3}{7}\selectfont, text=numEven] at (axis cs:1.127500000000000,1.03) {8};
\node[anchor=center, font=\fontsize{6.3}{7}\selectfont, text=numOdd]  at (axis cs:2.237500000000000,1.03) {9};
\node[anchor=center, font=\fontsize{6.3}{7}\selectfont, text=numEven] at (axis cs:2.867500000000000,1.03) {10};
\node[anchor=center, font=\fontsize{6.3}{7}\selectfont, text=numOdd]  at (axis cs:3.512499999999999,1.03) {11};
\node[anchor=center, font=\fontsize{6.3}{7}\selectfont, text=numEven] at (axis cs:4.802500000000000,1.03) {12};
\node[anchor=center, font=\fontsize{6.3}{7}\selectfont, text=numOdd]  at (axis cs:6.167500000000001,1.03) {13};
\node[anchor=center, font=\fontsize{6.3}{7}\selectfont, text=numEven] at (axis cs:7.982500000000000,1.03) {14};
\node[anchor=center, font=\fontsize{6.3}{7}\selectfont, text=numOdd]  at (axis cs:9.782500000000001,1.03) {15};
\node[anchor=center, font=\fontsize{6.3}{7}\selectfont, text=numEven] at (axis cs:12.812500000000000,1.03) {16};
\node[anchor=center, font=\fontsize{6.3}{7}\selectfont, text=numOdd]  at (axis cs:17.987500000000001,1.03) {17};
\node[anchor=center, font=\fontsize{6.3}{7}\selectfont, text=numEven] at (axis cs:24.992500000000000,1.03) {18};

\end{axis}

\end{tikzpicture}
  \caption{Sampling point distribution and ZF beamforming design \eqref{eq:Wzf} normalized by~\eqref{eq:column_power_normalization} for the line-source configuration in Fig.~\ref{fig:2D_line_source}a. In the upper plot, 14 beam locations are selected according to the DoF
estimate from  \eqref{eq:line_source_offset}. The numbered markers denote the beam focus. In the lower plot, the 18 number markers correspond to the DEIM-selected beam locations.}  \label{fig:2D_line_sampling_distribtuion}
\end{figure}
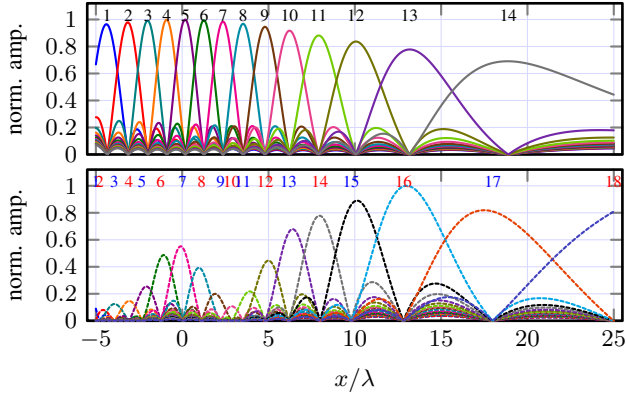
Figure~\ref{fig:2D_line_sampling_distribtuion} further presents the transverse
beam distribution obtained by superimposing the beamforming results at all
predefined sampling points. The upper plot shows the normalized ZF beam
patterns obtained from the 14 analytically predicted sampling points given by
\eqref{eq:line_source_offset}. After the column-power normalization in
\eqref{eq:column_power_normalization}, the field amplitudes represent the
relative focusing gains. As can be seen in Fig.~\ref{fig:sampling}, the representative focal positions are placed at the midpoints of the resolvable intervals, rather than at their boundaries. This midpoint placement leads to a more even cell-centered coverage of the receiving line.

The lower plot of Fig.~\ref{fig:2D_line_sampling_distribtuion} shows the ZF
beamforming result obtained with 18 DEIM-selected sampling points. Since the
number of prescribed beams exceeds the estimated DoF of 14, the ZF
beamforming problem becomes over-constrained. Each beam is required to focus
at its own target position while suppressing the field at the remaining target
points. For closely spaced targets, the corresponding channel vectors become
highly correlated, so the additional nulling constraints consume spatial DoF
that would otherwise contribute to focusing gain. As a result, some focal
points exhibit reduced main-lobe amplitudes, higher sidelobes, and stronger
overlap between neighboring beams. Along the transverse direction, the field
distribution still forms several localized peaks, each associated with one
selected sampling point, but their separability degrades when the number of
prescribed beams exceeds the available DoF.
\begin{figure}[t]
  \centering
\definecolor{c1}{named}{blue}
\definecolor{c2}{named}{red}
\definecolor{c3}{rgb}{0,0.5,0.5}
\definecolor{c4}{rgb}{1,0.45,0}
\definecolor{c5}{named}{violet}
\definecolor{c6}{rgb}{0,0.45,0}
\definecolor{c7}{named}{magenta}
\definecolor{c8}{rgb}{0,0.55,0.65}
\definecolor{c9}{rgb}{0.45,0.22,0.05}
\definecolor{c10}{rgb}{0.9,0.25,0.55}
\definecolor{c11}{rgb}{0.45,0.8,0}
\definecolor{c12}{rgb}{0.45,0.45,0}
\definecolor{c13}{rgb}{0.45,0.15,0.65}
\definecolor{c14}{rgb}{0.45,0.45,0.45}
\definecolor{c15}{named}{black}
\definecolor{c16}{rgb}{0,0.65,0.9}
\definecolor{c17}{rgb}{0.9,0.25,0}
\definecolor{c18}{rgb}{0.25,0.25,0.75}

\definecolor{numOdd}{named}{blue}
\definecolor{numEven}{named}{red}

\begin{tikzpicture}[
    line join=round,
    line cap=round,
    font=\fontsize{9}{11}\selectfont
]

\begin{axis}[
    name=topplot,
    at={(0cm,3.55cm)},
    anchor=south west,
    width=7.08cm,
    height=2cm,
    scale only axis,
    xmin=-5.5,
    xmax=25.5,
    ymin=0,
    ymax=1.12,
    xtick={-5,0,5,10,15,20,25},
    xticklabels={},
    ytick={0,0.2,0.4,0.6,0.8,1.0},
    ylabel={norm. amp.},
    xlabel={},
    ylabel style={font=\fontsize{9}{11}\selectfont},
    xticklabel style={font=\fontsize{9}{11}\selectfont},
    yticklabel style={font=\fontsize{9}{11}\selectfont},
    axis background/.style={fill=white},
    grid=major,
    major grid style={thin,blue!15!white},
    tick align=inside,
    line width=1pt,
    tick style={line width=0.8pt},
    clip=true
]

\addplot[c1,  line width=0.8pt] table[x index=0, y index=1]  {data/beam_width.dat};
\addplot[c2,  line width=0.8pt] table[x index=0, y index=2]  {data/beam_width.dat};
\addplot[c3,  line width=0.8pt] table[x index=0, y index=3]  {data/beam_width.dat};
\addplot[c4,  line width=0.8pt] table[x index=0, y index=4]  {data/beam_width.dat};
\addplot[c5,  line width=0.8pt] table[x index=0, y index=5]  {data/beam_width.dat};
\addplot[c6,  line width=0.8pt] table[x index=0, y index=6]  {data/beam_width.dat};
\addplot[c7,  line width=0.8pt] table[x index=0, y index=7]  {data/beam_width.dat};
\addplot[c8,  line width=0.8pt] table[x index=0, y index=8]  {data/beam_width.dat};
\addplot[c9,  line width=0.8pt] table[x index=0, y index=9]  {data/beam_width.dat};
\addplot[c10, line width=0.8pt] table[x index=0, y index=10] {data/beam_width.dat};
\addplot[c11, line width=0.8pt] table[x index=0, y index=11] {data/beam_width.dat};
\addplot[c12, line width=0.8pt] table[x index=0, y index=12] {data/beam_width.dat};
\addplot[c13, line width=0.8pt] table[x index=0, y index=13] {data/beam_width.dat};
\addplot[c14, line width=0.8pt] table[x index=0, y index=14] {data/beam_width.dat};

\node[anchor=center, font=\fontsize{6.5}{7}\selectfont, text=black] at (axis cs:-4.342356286919999,1.03) {1};
\node[anchor=center, font=\fontsize{6.5}{7}\selectfont, text=black] at (axis cs:-3.119510095320000,1.03) {2};
\node[anchor=center, font=\fontsize{6.5}{7}\selectfont, text=black] at (axis cs:-1.979415354400000,1.03) {3};
\node[anchor=center, font=\fontsize{6.5}{7}\selectfont, text=black] at (axis cs:-0.886676660540000,1.03) {4};
\node[anchor=center, font=\fontsize{6.5}{7}\selectfont, text=black] at (axis cs:0.186025300580000,1.03) {5};
\node[anchor=center, font=\fontsize{6.5}{7}\selectfont, text=black] at (axis cs:1.262873195260000,1.03) {6};
\node[anchor=center, font=\fontsize{6.5}{7}\selectfont, text=black] at (axis cs:2.368685231880000,1.03) {7};
\node[anchor=center, font=\fontsize{6.5}{7}\selectfont, text=black] at (axis cs:3.532916472260000,1.03) {8};
\node[anchor=center, font=\fontsize{6.5}{7}\selectfont, text=black] at (axis cs:4.795444418360000,1.03) {9};
\node[anchor=center, font=\fontsize{6.5}{7}\selectfont, text=black] at (axis cs:6.217049693320000,1.03) {10};
\node[anchor=center, font=\fontsize{6.5}{7}\selectfont, text=black] at (axis cs:7.901915102559999,1.03) {11};
\node[anchor=center, font=\fontsize{6.5}{7}\selectfont, text=black] at (axis cs:10.054957102219998,1.03) {12};
\node[anchor=center, font=\fontsize{6.5}{7}\selectfont, text=black] at (axis cs:13.166002301019999,1.03) {13};
\node[anchor=center, font=\fontsize{6.5}{7}\selectfont, text=black] at (axis cs:18.891239312019998,1.03) {14};

\end{axis}

\begin{axis}[
    name=bottomplot,
    at={(0cm,1.35cm)},
    anchor=south west,
    width=7.08cm,
    height=2cm,
    scale only axis,
    xmin=-5.5,
    xmax=25.5,
    ymin=0,
    ymax=1.12,
    xtick={-5,0,5,10,15,20,25},
    ytick={0,0.2,0.4,0.6,0.8,1.0},
    ylabel={norm. amp.},
    xlabel={$x/\lambda$},
    xlabel style={font=\fontsize{9}{11}\selectfont},
    ylabel style={font=\fontsize{9}{11}\selectfont},
    xticklabel style={font=\fontsize{9}{11}\selectfont},
    yticklabel style={font=\fontsize{9}{11}\selectfont},
    axis background/.style={fill=white},
    grid=major,
    major grid style={thin,blue!15!white},
    tick align=inside,
    line width=1pt,
    tick style={line width=0.8pt},
    clip=false
]

\addplot[c1,  dash pattern=on 1.2pt off 1.0pt, line width=0.8pt] table[x index=0, y index=1]  {data/beam_width_DEIM.dat};
\addplot[c2,  dash pattern=on 1.2pt off 1.0pt, line width=0.8pt] table[x index=0, y index=2]  {data/beam_width_DEIM.dat};
\addplot[c3,  dash pattern=on 1.2pt off 1.0pt, line width=0.8pt] table[x index=0, y index=3]  {data/beam_width_DEIM.dat};
\addplot[c4,  dash pattern=on 1.2pt off 1.0pt, line width=0.8pt] table[x index=0, y index=4]  {data/beam_width_DEIM.dat};
\addplot[c5,  dash pattern=on 1.2pt off 1.0pt, line width=0.8pt] table[x index=0, y index=5]  {data/beam_width_DEIM.dat};
\addplot[c6,  dash pattern=on 1.2pt off 1.0pt, line width=0.8pt] table[x index=0, y index=6]  {data/beam_width_DEIM.dat};
\addplot[c7,  dash pattern=on 1.2pt off 1.0pt, line width=0.8pt] table[x index=0, y index=7]  {data/beam_width_DEIM.dat};
\addplot[c8,  dash pattern=on 1.2pt off 1.0pt, line width=0.8pt] table[x index=0, y index=8]  {data/beam_width_DEIM.dat};
\addplot[c9,  dash pattern=on 1.2pt off 1.0pt, line width=0.8pt] table[x index=0, y index=9]  {data/beam_width_DEIM.dat};
\addplot[c10, dash pattern=on 1.2pt off 1.0pt, line width=0.8pt] table[x index=0, y index=10] {data/beam_width_DEIM.dat};
\addplot[c11, dash pattern=on 1.2pt off 1.0pt, line width=0.8pt] table[x index=0, y index=11] {data/beam_width_DEIM.dat};
\addplot[c12, dash pattern=on 1.2pt off 1.0pt, line width=0.8pt] table[x index=0, y index=12] {data/beam_width_DEIM.dat};
\addplot[c13, dash pattern=on 1.2pt off 1.0pt, line width=0.8pt] table[x index=0, y index=13] {data/beam_width_DEIM.dat};
\addplot[c14, dash pattern=on 1.2pt off 1.0pt, line width=0.8pt] table[x index=0, y index=14] {data/beam_width_DEIM.dat};
\addplot[c15, dash pattern=on 1.2pt off 1.0pt, line width=0.8pt] table[x index=0, y index=15] {data/beam_width_DEIM.dat};
\addplot[c16, dash pattern=on 1.2pt off 1.0pt, line width=0.8pt] table[x index=0, y index=16] {data/beam_width_DEIM.dat};
\addplot[c17, dash pattern=on 1.2pt off 1.0pt, line width=0.8pt] table[x index=0, y index=17] {data/beam_width_DEIM.dat};
\addplot[c18, dash pattern=on 1.2pt off 1.0pt, line width=0.8pt] table[x index=0, y index=18] {data/beam_width_DEIM.dat};

\node[anchor=center, font=\fontsize{6.3}{7}\selectfont, text=numOdd]  at (axis cs:-4.992500000000000,1.03) {1};
\node[anchor=center, font=\fontsize{6.3}{7}\selectfont, text=numEven] at (axis cs:-4.782500000000000,1.03) {2};
\node[anchor=center, font=\fontsize{6.3}{7}\selectfont, text=numOdd]  at (axis cs:-3.927500000000000,1.03) {3};
\node[anchor=center, font=\fontsize{6.3}{7}\selectfont, text=numEven] at (axis cs:-3.087500000000000,1.03) {4};
\node[anchor=center, font=\fontsize{6.3}{7}\selectfont, text=numOdd]  at (axis cs:-2.322500000000000,1.03) {5};
\node[anchor=center, font=\fontsize{6.3}{7}\selectfont, text=numEven] at (axis cs:-1.242500000000000,1.03) {6};
\node[anchor=center, font=\fontsize{6.3}{7}\selectfont, text=numOdd]  at (axis cs:0.032500000000001,1.03) {7};
\node[anchor=center, font=\fontsize{6.3}{7}\selectfont, text=numEven] at (axis cs:1.127500000000000,1.03) {8};
\node[anchor=center, font=\fontsize{6.3}{7}\selectfont, text=numOdd]  at (axis cs:2.237500000000000,1.03) {9};
\node[anchor=center, font=\fontsize{6.3}{7}\selectfont, text=numEven] at (axis cs:2.867500000000000,1.03) {10};
\node[anchor=center, font=\fontsize{6.3}{7}\selectfont, text=numOdd]  at (axis cs:3.512499999999999,1.03) {11};
\node[anchor=center, font=\fontsize{6.3}{7}\selectfont, text=numEven] at (axis cs:4.802500000000000,1.03) {12};
\node[anchor=center, font=\fontsize{6.3}{7}\selectfont, text=numOdd]  at (axis cs:6.167500000000001,1.03) {13};
\node[anchor=center, font=\fontsize{6.3}{7}\selectfont, text=numEven] at (axis cs:7.982500000000000,1.03) {14};
\node[anchor=center, font=\fontsize{6.3}{7}\selectfont, text=numOdd]  at (axis cs:9.782500000000001,1.03) {15};
\node[anchor=center, font=\fontsize{6.3}{7}\selectfont, text=numEven] at (axis cs:12.812500000000000,1.03) {16};
\node[anchor=center, font=\fontsize{6.3}{7}\selectfont, text=numOdd]  at (axis cs:17.987500000000001,1.03) {17};
\node[anchor=center, font=\fontsize{6.3}{7}\selectfont, text=numEven] at (axis cs:24.992500000000000,1.03) {18};

\end{axis}

\end{tikzpicture}
  \caption{Sampling point distribution and normalized MRT beamforming design \eqref{eq:mrt_weight} normalized by~\eqref{eq:column_power_normalization} for the line-source configuration in Fig.~\ref{fig:2D_line_source}a, using the same geometric
parameters as in Fig~\ref{fig:2D_line_sampling_distribtuion}. The numbered markers indicate the 14 theoretically predicted beam locations in the upper plot, and the 18 DEIM-selected beam locations in the lower plot.}  \label{fig:2D_line_sampling_distribtuion_MRT}
\end{figure}
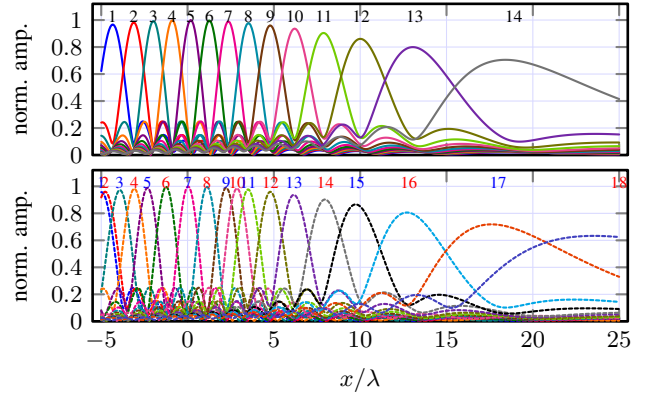

In the upper plot of Fig.~\ref{fig:2D_line_sampling_distribtuion_MRT} for MRT beamforming, 14 beams are synthesized using the analytically predicted sampling points, which exactly match the maximum DoF. The obtained beam distribution is similar to that of the ZF scheme.
The resulting beams remain well organized and distinguishable, even though adjacent beams partially overlap. Moreover, adjacent beams tend to produce field nulls at the peak location of the beam under
consideration. In the lower plot, however, 18 beams are synthesized, which exceeds 14. Consequently, some neighboring beams overlap severely, with the overlapping
regions far exceeding the $1/\sqrt{2}$ amplitude level. Several focal peaks
can therefore no longer be effectively identified as distinct beams. This
indicates that exceeding the DoF limit significantly degrades beam
separability and prevents the formation of fully independent focal spots.

\begin{figure}[t]
  \centering
\includegraphics[width=\linewidth]{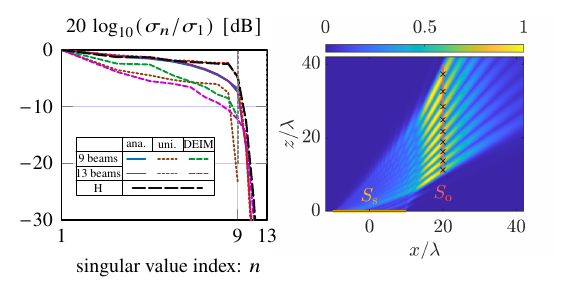}
  \caption{Normalized singular values and beamforming intensity distribution for the line-source configuration in Fig.~\ref{fig:2D_line_source}b. The left panel compares 9, 13 beams using analytical, uniform sampling and the DEIM method. \(\M H\) denotes the result obtained from the continuous source and observation regions. 
The right plot shows the normalized electric-field distribution for 9 beams in the \(x\)-\(z\) plane using MRT  beamforming~\eqref{eq:mrt_weight}. }  \label{fig:2D_line_sampling_distribtuion_depth}
\end{figure}
For the longitudinal beamforming case, we consider the line-source
configuration in Fig.~\ref{fig:2D_line_source}b. The verification parameters
are chosen as \(d_1=d_2=\ell_1=\ell_2/3=10\lambda\). The resulting modal
behavior is first examined through the singular-value spectra in
Fig.~\ref{fig:2D_line_sampling_distribtuion_depth}. The analytically selected
9 sampling points are consistent with the longitudinal DoF estimate~\eqref{eq:LTR_perp}. When 13 beams are prescribed, the singular-value spectrum exceeds the dominant modal region, indicating that the additional constraints are weakly supported by the channel.

For both the uniform and DEIM-based sampling schemes, the singular-value spectra exhibit an earlier knee at the 8th DoF, whereas the proposed analytical method preserves the knee at the 9th DoF. Although the uniform scheme gives a relatively flat spectrum, most of its unnormalized dominant singular values are smaller than those of the analytical method. This further demonstrates the advantage of the proposed sampling strategy.

The corresponding 2D field distribution in
Fig.~\ref{fig:2D_line_sampling_distribtuion_depth} further illustrates the
spatial focusing behavior in the \(x\)-\(z\) plane. The prescribed focal
locations lie along the vertical observation line. The field intensity forms
localized high-field regions around these locations, while the beam energy
spreads gradually as the propagation distance increases. Since the whole source
region \(S_{\mathrm{s}}\) contributes to the radiation, the field pattern is
formed by the superposition of the focused responses from the entire aperture.

\begin{figure}[t]
  \centering
\definecolor{c1}{named}{blue}
\definecolor{c2}{named}{red}
\definecolor{c3}{rgb}{0,0.5,0.5}
\definecolor{c4}{rgb}{1,0.45,0}
\definecolor{c5}{named}{violet}
\definecolor{c6}{rgb}{0.55,0.34,0.29}
\definecolor{c7}{rgb}{0.89,0.47,0.76}
\definecolor{c8}{rgb}{0.25,0.25,0.25}
\definecolor{c9}{rgb}{0.00,0.60,0.20}

\def\pA{11.228203995035098}
\def\pB{13.674583490321529}
\def\pC{16.189336288250534}
\def\pD{18.848533497114907}
\def\pE{21.723089505726122}
\def\pF{24.891281957884875}
\def\pG{28.449061076929979}
\def\pH{32.522389118021522}
\def\pI{37.285377830569452}

\begin{tikzpicture}[
    line join=round,
    line cap=round,
    font=\fontsize{9}{11}\selectfont
]

\begin{axis}[
    name=upperplot,
    at={(0cm,2.20cm)},
    anchor=south west,
    width=7.08cm,
    height=2cm,
    scale only axis,
    xmin=10,
    xmax=40,
    ymin=0,
    ymax=1.12,
    xtick={10,15,20,25,30,35,40},
    xticklabels={},
    ytick={0,0.2,0.4,0.6,0.8,1.0},
    ylabel={norm. amp.},
    xlabel={},
    xlabel style={font=\fontsize{9}{11}\selectfont},
    ylabel style={font=\fontsize{9}{11}\selectfont},
    xticklabel style={font=\fontsize{9}{11}\selectfont},
    yticklabel style={font=\fontsize{9}{11}\selectfont},
    axis background/.style={fill=white},
    grid=major,
    major grid style={thin,blue!15!white},
    tick align=inside,
    line width=1pt,
    tick style={line width=0.8pt},
    clip=false
]

\addplot[c1, line width=0.8pt]
    table[x index=0, y index=1] {data/beam_depth_ZF.dat};

\addplot[c2, line width=0.8pt]
    table[x index=0, y index=2] {data/beam_depth_ZF.dat};

\addplot[c3, line width=0.8pt]
    table[x index=0, y index=3] {data/beam_depth_ZF.dat};

\addplot[c4, line width=0.8pt]
    table[x index=0, y index=4] {data/beam_depth_ZF.dat};

\addplot[c5, line width=0.8pt]
    table[x index=0, y index=5] {data/beam_depth_ZF.dat};

\addplot[c6, line width=0.8pt]
    table[x index=0, y index=6] {data/beam_depth_ZF.dat};

\addplot[c7, line width=0.8pt]
    table[x index=0, y index=7] {data/beam_depth_ZF.dat};

\addplot[c8, line width=0.8pt]
    table[x index=0, y index=8] {data/beam_depth_ZF.dat};

\addplot[c9, line width=0.8pt]
    table[x index=0, y index=9] {data/beam_depth_ZF.dat};

\node[
    anchor=north east,
    font=\fontsize{8.5}{10}\selectfont,
    fill=white,
    fill opacity=0.85,
    text opacity=1,
    inner sep=1.2pt
] at (rel axis cs:0.98,0.9) {ZF};

\node[anchor=center, font=\fontsize{6.5}{7}\selectfont, text=black] at (axis cs:\pA,1.03) {1};
\node[anchor=center, font=\fontsize{6.5}{7}\selectfont, text=black] at (axis cs:\pB,1.03) {2};
\node[anchor=center, font=\fontsize{6.5}{7}\selectfont, text=black] at (axis cs:\pC,1.03) {3};
\node[anchor=center, font=\fontsize{6.5}{7}\selectfont, text=black] at (axis cs:\pD,1.03) {4};
\node[anchor=center, font=\fontsize{6.5}{7}\selectfont, text=black] at (axis cs:\pE,1.03) {5};
\node[anchor=center, font=\fontsize{6.5}{7}\selectfont, text=black] at (axis cs:\pF,1.03) {6};
\node[anchor=center, font=\fontsize{6.5}{7}\selectfont, text=black] at (axis cs:\pG,1.03) {7};
\node[anchor=center, font=\fontsize{6.5}{7}\selectfont, text=black] at (axis cs:\pH,1.03) {8};
\node[anchor=center, font=\fontsize{6.5}{7}\selectfont, text=black] at (axis cs:\pI,1.03) {9};

\end{axis}

\begin{axis}[
    name=lowerplot,
    at={(0cm,0cm)},
    anchor=south west,
    width=7.08cm,
    height=2cm,
    scale only axis,
    xmin=10,
    xmax=40,
    ymin=0,
    ymax=1.12,
    xtick={10,15,20,25,30,35,40},
    ytick={0,0.2,0.4,0.6,0.8,1.0},
    ylabel={norm. amp.},
    xlabel={$z/\lambda$},
    xlabel style={font=\fontsize{9}{11}\selectfont},
    ylabel style={font=\fontsize{9}{11}\selectfont},
    xticklabel style={font=\fontsize{9}{11}\selectfont},
    yticklabel style={font=\fontsize{9}{11}\selectfont},
    axis background/.style={fill=white},
    grid=major,
    major grid style={thin,blue!15!white},
    tick align=inside,
    line width=1pt,
    tick style={line width=0.8pt},
    clip=false
]

\addplot[c1, dash pattern=on 1.2pt off 1.0pt, line width=0.8pt]
    table[x index=0, y index=1] {data/beam_depth.dat};

\addplot[c2, dash pattern=on 1.2pt off 1.0pt, line width=0.8pt]
    table[x index=0, y index=2] {data/beam_depth.dat};

\addplot[c3, dash pattern=on 1.2pt off 1.0pt, line width=0.8pt]
    table[x index=0, y index=3] {data/beam_depth.dat};

\addplot[c4, dash pattern=on 1.2pt off 1.0pt, line width=0.8pt]
    table[x index=0, y index=4] {data/beam_depth.dat};

\addplot[c5, dash pattern=on 1.2pt off 1.0pt, line width=0.8pt]
    table[x index=0, y index=5] {data/beam_depth.dat};

\addplot[c6, dash pattern=on 1.2pt off 1.0pt, line width=0.8pt]
    table[x index=0, y index=6] {data/beam_depth.dat};

\addplot[c7, dash pattern=on 1.2pt off 1.0pt, line width=0.8pt]
    table[x index=0, y index=7] {data/beam_depth.dat};

\addplot[c8, dash pattern=on 1.2pt off 1.0pt, line width=0.8pt]
    table[x index=0, y index=8] {data/beam_depth.dat};

\addplot[c9, dash pattern=on 1.2pt off 1.0pt, line width=0.8pt]
    table[x index=0, y index=9] {data/beam_depth.dat};

\node[
    anchor=north east,
    font=\fontsize{8.5}{10}\selectfont,
    fill=white,
    fill opacity=0.85,
    text opacity=1,
    inner sep=1.2pt
] at (rel axis cs:0.98,0.9) {MRT};

\node[anchor=center, font=\fontsize{6.5}{7}\selectfont, text=black] at (axis cs:\pA,1.03) {1};
\node[anchor=center, font=\fontsize{6.5}{7}\selectfont, text=black] at (axis cs:\pB,1.03) {2};
\node[anchor=center, font=\fontsize{6.5}{7}\selectfont, text=black] at (axis cs:\pC,1.03) {3};
\node[anchor=center, font=\fontsize{6.5}{7}\selectfont, text=black] at (axis cs:\pD,1.03) {4};
\node[anchor=center, font=\fontsize{6.5}{7}\selectfont, text=black] at (axis cs:\pE,1.03) {5};
\node[anchor=center, font=\fontsize{6.5}{7}\selectfont, text=black] at (axis cs:\pF,1.03) {6};
\node[anchor=center, font=\fontsize{6.5}{7}\selectfont, text=black] at (axis cs:\pG,1.03) {7};
\node[anchor=center, font=\fontsize{6.5}{7}\selectfont, text=black] at (axis cs:\pH,1.03) {8};
\node[anchor=center, font=\fontsize{6.5}{7}\selectfont, text=black] at (axis cs:\pI,1.03) {9};

\end{axis}

\end{tikzpicture}
  \caption{Sampling-point distribution and beamforming performance normalized by~\eqref{eq:column_power_normalization} for the line-source configuration in Fig.~\ref{fig:2D_line_source}b.
The two main panels show the normalized beam amplitudes along the vertical observation line for ZF~\eqref{eq:Wzf} and MRT~\eqref{eq:mrt_weight} beamforming, respectively. 
The numbered markers indicate the focal locations selected according to
the longitudinal DoF. }  \label{fig:beamdepth}
\end{figure}
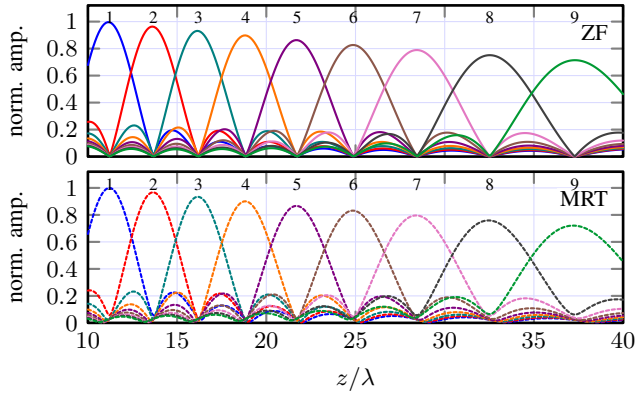
Figure~\ref{fig:beamdepth} then gives a more
direct comparison of the beam amplitudes along the vertical observation line.
The sampling positions are determined from the closed-form DoF-based~\eqref{eq:LTR_perp} sampling
rule, with each sample placed at the midpoint of its corresponding resolvable
interval rather than directly on the boundary.

For both ZF and MRT beamforming, the generated beams are centered around the
predicted focal positions, showing that the longitudinal DoF-based sampling
rule can guide the placement of independently controllable focal regions.
Compared with MRT, the ZF method provides sharper separation between adjacent
focal points and lower leakage at the other target locations, because it
explicitly imposes nulling constraints among different sampling points. In
contrast, MRT mainly maximizes the response at each desired focal point and
therefore exhibits stronger overlap and higher sidelobe levels.

 These results confirm that the estimated DoF not only predicts the number of resolvable longitudinal regions, but also provides a practical basis for beam synthesis along the receiving line. When the number of prescribed beams remains close to the available DoF, the focal peaks remain distinguishable, with adjacent-beam overlap approximately bounded by the $1/\sqrt{2}$ field-amplitude level. Prescribing more beams than the dominant modal support would increase the overlap between neighboring beams and reduce beam separability, especially in the farther region where the field variation becomes smoother.

\section{3D Surface Source}\label{sec:3D Surface Source}

\begin{figure}[t]
  \centering  \includegraphics[width=0.9\linewidth]{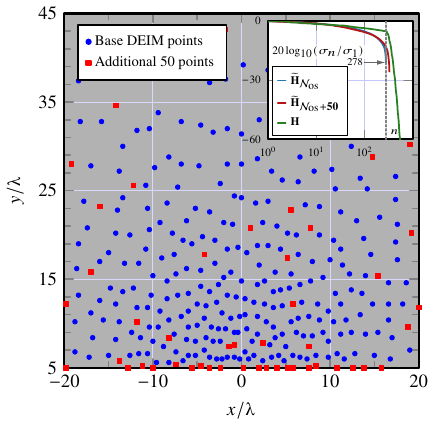}
  \caption{DEIM-selected sampling points on the candidate grid in the $x$--$y$ plane on the floor (see inset in Fig.~\ref{fig:numerical_closed_dof}), $d_1=d_2=5\lambda$, $l_1=w_1=20\lambda, l_2=w_2=40\lambda$. To demonstrate the effectiveness of the DEIM method, an additional 50 sampling points are introduced based on the theoretical DoF results \eqref{eq:area_LIS_floor}. The inset compares the normalized singular values of three propagation matrices:
$\tM{H}_{\mathcal{N}_{\T{os}}}$ constructed from the DEIM-selected sampling points with the number of points equal to the DoF.
$\tM{H}_{\mathcal{N}_{\T{os}}+50}$ constructed by adding 50 DEIM-selected points. 
$\M{H}$ corresponding to the equidistantly sampled reference channel. Some sampling points are partially obscured by the inset, see also Fig.~\ref{fig:3D_surface_sampling_distribtuion}.}  \label{fig:DEIM_points}
\end{figure}

For LIS applications, a particularly relevant configuration is a surface
aperture deployed on a wall to serve users or devices over a finite observation
region. Therefore, we focus on this representative configuration, as shown in Fig.~\ref{fig:numerical_closed_dof}.
 Specifically, a dense candidate grid is first generated on the $x$--$y$ plane, and the channel matrix evaluated from the Green's function \eqref{eq:transmission matrices} from the discretized LIS to all candidate points is constructed. The theoretical number of DoF ${\mathcal{N}}_{\T{os}}$ is obtained from the mutual shadow area \eqref{eq:area_LIS_floor}.
Accordingly, the first left singular vectors are retained, and the standard DEIM greedy procedure is then applied to identify the interpolation points.

Fig.~\ref{fig:DEIM_points} illustrates the DEIM-selected sampling points on the candidate grid. Most of the points are concentrated in the region close to the LIS, while their density gradually decreases as the distance from the LIS increases. This pattern suggests that the field varies more rapidly in the near-LIS region, where a denser set of sampling points is needed to capture the dominant spatial modes accurately. In contrast, fewer points are sufficient in regions farther away. Moreover, after adding 50 extra sampling points on top of the original DEIM-selected set, the knee point of the resulting singular value curve becomes very close to that obtained from the full transmission matrix \eqref{eq:transmission matrices}. This indicates that the sampling points selected by the DEIM algorithm can effectively characterize the DoF within the observation region.

The inset compares the normalized singular value distributions for three propagation matrices constructed with different sampling sets. 
It can be observed that the three curves exhibit a similar singular value decay trend and show a clear drop after the DoF knee point. 
This indicates that the base DEIM-selected points already capture the dominant spatial modes of the propagation field. 
The additional 50 points only slightly modify the tail of the singular value distribution and contribute little to the DoF.

 \begin{figure}[t]
  \centering
\includegraphics[width=1\linewidth]{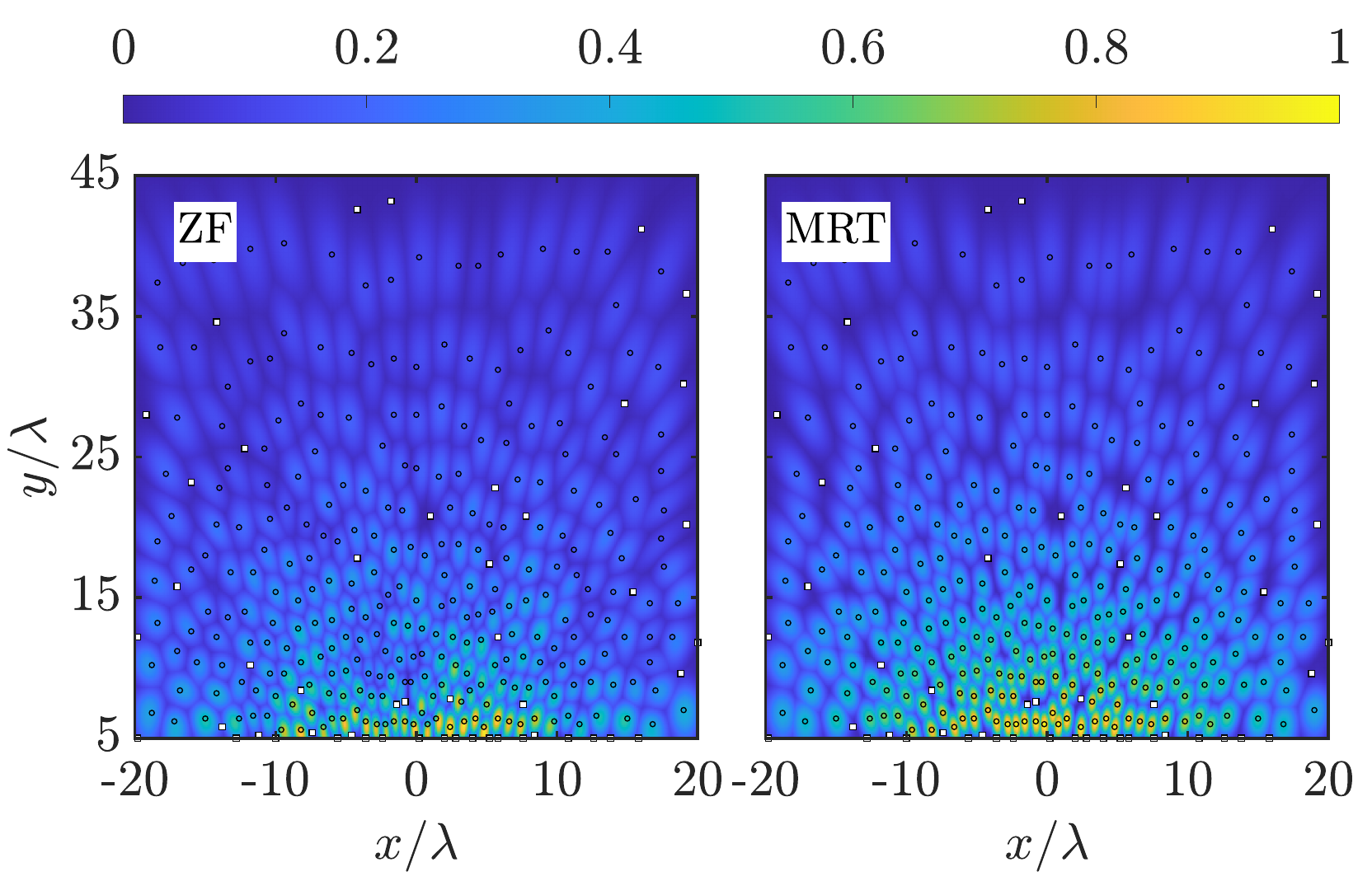}
  \caption{Normalized beamforming intensity map using ZF and MRT methods in the $x$--$y$ plane obtained by focusing on the DEIM-selected sampling points. The black hollow circles denote the sampling points selected by the DEIM algorithm with beamforming, while the white square markers indicate the locations of the additional 50 sampling points without beamforming. The simulation parameters are the same as those used in Fig.~\ref{fig:DEIM_points}.}  \label{fig:3D_surface_sampling_distribtuion}
\end{figure}

To further evaluate the effectiveness of the selected points, MRT and ZF beamforming are performed toward each DEIM location, and the resulting field intensity distribution is computed on the $x$--$y$ plane. The corresponding beamforming map is shown in Fig.~\ref{fig:3D_surface_sampling_distribtuion}. The ZF and MRT methods exhibit very similar beamforming distributions. However, in the ZF method, some beam peaks deviate from the prescribed focal points in order to improve the isolation from other beams. The high-intensity regions are mainly concentrated near the DEIM-selected locations and gradually spread across the observation region through overlapping focal patterns. This result confirms that the selected points are closely related to the dominant spatial modes of the propagation operator and are sufficient to characterize the main energy distribution of the 3D surface-source channel.

The above results demonstrate that, for the considered 3D surface-source
geometry, the DEIM-based sampling strategy can efficiently select
representative spatial points from the propagation matrix. The number of
selected points agrees well with the theoretical DoF predicted by the mutual
shadow area formula, and the resulting beamforming patterns show that these
points capture the dominant field structure in the observation region.
Therefore, the proposed mutual-shadow-area and DEIM framework provides a
physically interpretable and computationally efficient approach for sampling
design and reduced-order modeling in near-field 3D planar-source scenarios.

\section{Polarization-Aware Beamforming}\label{sec:Polarization-Aware Beamforming}

Section~\ref{sec:3D Surface Source} shows that the DEIM-based strategy can effectively identify
representative sampling locations for a 3D surface-source channel. However,
that analysis mainly considers the spatial field distribution without
explicitly separating the polarization components. Since practical near-field
DoF and received power depend on both spatial variation and field
polarization, we further investigate the impact of polarization on the
sampling distribution.

For a polarization-aware formulation, the propagation operator is extended from a scalar transmission matrix to a vector-valued transmission matrix. Specifically, for each candidate observation point on the $x$--$y$ plane, the electric field generated by the discretized LIS is decomposed into its three Cartesian components, i.e.,
$
\V{E}(\V{r})
=
E_\T x(\V{r})\UV x
+
E_\T y(\V{r})\UV y
+
E_\T z(\V{r})\UV z .
$ The channel matrix and Green's function \eqref{eq:dyadic_eq} is applied.
Compared with the scalar formulation, this augmented matrix jointly characterizes the spatial and polarization-domain responses of the 3D surface-source channel.

\subsection{Polarization Effects on DoF and sampling}
\begin{figure}[t]
  \centering  \includegraphics[width=\linewidth]{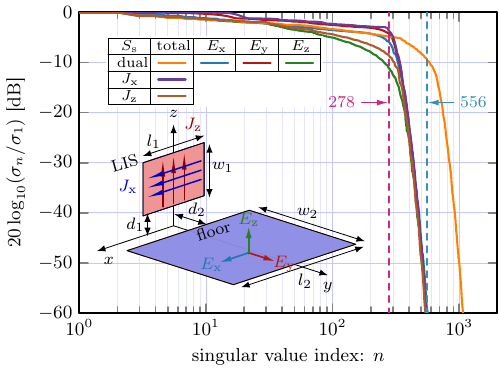}
  \caption{Normalized singular values $\sigma_n/\sigma_1$ for the LIS and the floor on a dB scale, under the same geometric setting as in Fig.~\ref{fig:DEIM_points}. For the dual-polarized configuration, both $J_{\T{x}}$ and $J_{\T{z}}$ currents are excited, and the singular value distributions are shown for the resulting total electric field as well as its $E_\T x$, $E_\T y$, and $E_\T z$ components. 
For the single-polarized configurations, the singular values are obtained from the total electric field generated by exciting either the $J_{\T{x}}$ current or the $J_{\T{z}}$ current alone. 
The red and blue dashed lines correspond to $\mathcal{N}_{\T{os}}=278$ for a single polarization and $556$ for dual polarizations from number of DoF \eqref{eq:line_source_short} and mutual shadow area \eqref{eq:area_LIS_floor}, \eqref{eq:area_LIS_floor_2}, respectively.}
  \label{fig:numerical_closed_dof_polarization}
\end{figure}

For the near-field application scenario in
Fig.~\ref{fig:numerical_closed_dof}, we examine how the three polarization
components contribute to the spatial DoF. To this end, we compare
the singular value spectrum of the total field with those of the individual field
components. 

When the transmitted current is dual-polarized (see Fig.~\ref{fig:numerical_closed_dof_polarization}), more observable modes are excited than in the single-polarized configurations. This can be seen from the delayed decay of the dual-polarized total-field spectrum compared with the spectra generated by the individual current polarizations. In the present configuration, the dual-polarized configuration approximately doubles the effective modal range relative to the single-polarized configuration, which agrees with the theoretical prediction \eqref{eq:dof(7)}. However, this increase comes from the complementary field responses generated
by different source polarizations, rather than from duplicating identical
spatial modes.

The individual Cartesian components also exhibit different modal behaviors. For the dual-polarized electric current, the normalized singular values for $E_{\T{z}}$ component decays faster than the $E_{\T{x}}$ and $E_{\T{y}}$ components, while $E_{\T{x}}$ and $E_{\T{y}}$ maintain relatively large normalized singular values over a wider range of modal indices. This indicates that the considered geometry and polarization configuration lead to an imbalance among the observable Cartesian field components. The field components therefore do not contribute equally to the effective near-field modal space.

Finally, comparing the component-wise spectra with the total-field spectrum shows that jointly observing $E_{\T{x}}$, $E_{\T{y}}$, and $E_{\T{z}}$ provides more effective modes than using a single field component alone. Nevertheless, the total-field spectrum is not a simple sum of the spectra of the three Cartesian components, and the corresponding DoF does not increase by a factor of three. This is because the three electric-field components are coupled through the underlying electromagnetic propagation mechanism and therefore contain partially overlapping spatial information. Similarly, different current polarizations enrich the modal space, but their contributions are not completely independent. Thus, Fig.~\ref{fig:numerical_closed_dof_polarization} shows that
multi-polarization transmission increases the observable near-field DoF, but
only to two polarization DoF.

Building on this observation, we further examine whether introducing additional
source types can provide extra DoF. Besides the dual-polarized electric currents
considered above, we also include, or separately excite, dual-polarized magnetic
currents on the source surface (see $\M{G_\T J}$ and $\M{G_\T M}$ in App.~\ref{app:2D_Source}). The resulting singular-value spectra show that
magnetic-current sources do not further increase the number of DoF for the
considered surface-based LoS configuration. This is because the electric and
magnetic equivalent currents excite the same set of propagating electromagnetic
modes, although they may redistribute the modal energy among different field
components. Hence, the available DoF are determined by the physical propagation
operator between the source and observation regions, rather than by the number
of equivalent source-current components.

\subsection{Polarization-Aware Sampling and Beamforming}
\begin{figure}[t]
  \centering  \definecolor{myblue}{rgb}{0.12157,0.47059,0.70588}
\definecolor{myred}{rgb}{0.71373,0.08235,0.08627}
\definecolor{mygreen}{rgb}{0.16078,0.50196,0.13725}

\begin{tikzpicture}[
line join=round,
line cap=round,
font=\fontsize{9}{11}\selectfont
]

\begin{axis}[
name=mainplot,
width=6cm,
height=6cm,
scale only axis,
xmin=-20,
xmax=20,
ymin=5,
ymax=45,
xtick={-20,-10,0,10,20},
ytick={5,15,25,35,45},
minor x tick num=1,
minor y tick num=4,
xlabel={$x/\lambda$},
ylabel={$y/\lambda$},
xlabel style={font=\fontsize{9}{11}\selectfont},
ylabel style={font=\fontsize{9}{11}\selectfont},
title style={font=\fontsize{9}{11}\selectfont},
xticklabel style={font=\fontsize{9}{11}\selectfont},
yticklabel style={font=\fontsize{9}{11}\selectfont},
axis lines=box,
axis line style={black, line width=1pt},
tick align=inside,
tick style={line width=0.8pt},
major grid style={thin,blue!15!white},
minor grid style={thin,blue!8!white},
grid=major,
axis background/.style={fill=white},
clip=false,
axis on top,
unbounded coords=discard,
legend style={
    at={(0.99,0.36)},
    anchor=north east,
    fill=white,
    draw=black,
    line width=0.8pt,
    font=\fontsize{7}{10}\selectfont,
    cells={anchor=west},
    inner xsep=4pt,
    inner ysep=3pt
},
legend cell align=left
]

\fill[black!30] (axis cs:-20,5) rectangle (axis cs:20,45);


\addplot[
    only marks,
    color=myred,
    mark=square*,
    mark size=1.2pt,
    mark options={fill=myred},
]
table[
    x expr={-\thisrowno{3}},
    y index=2
] {data/base_deim_points_polarization.tsv};

\addplot[
    only marks,
    color=myblue,
    mark=*,
    mark size=1.2pt,
    mark options={fill=myblue},
]
table[
    x expr={-\thisrowno{1}},
    y index=0
] {data/base_deim_points_polarization.tsv};

\addplot[
    only marks,
    color=mygreen,
    mark=diamond*,
    mark size=1.5pt,
    mark options={fill=mygreen},
]
table[
    x expr={-\thisrowno{5}},
    y index=4
] {data/base_deim_points_polarization.tsv};

\legend{$E_\mathrm{x}$, $E_\mathrm{y}$, $E_\mathrm{z}$}

\end{axis}

\end{tikzpicture}
  \caption{DEIM-selected sampling points on the candidate grid in the $x$--$y$ plane on the floor for dual-polarized current on the LIS, under the same conditions as in Fig.~\ref{fig:DEIM_points}. }  \label{fig:DEIM_points_polarization}
\end{figure}
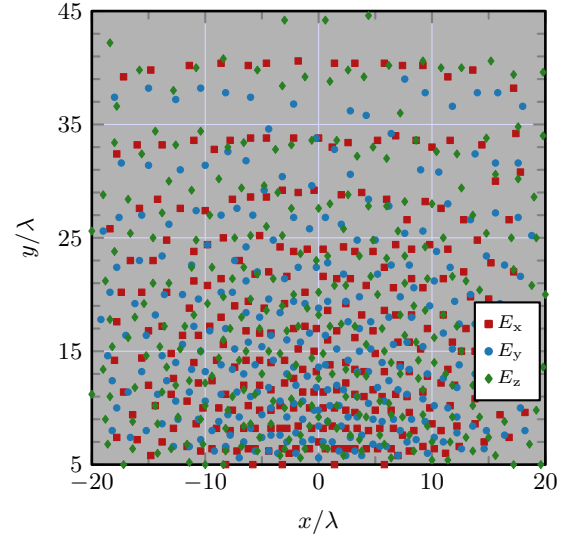

Figure~\ref{fig:DEIM_points_polarization} shows the DEIM-selected sampling points on the candidate observation grid in the \(x\)-\(y\) plane for the dual-polarized current excitation setup in Fig.~\ref{fig:numerical_closed_dof_polarization}. In this case, \(E_{\T{x}}\) and \(E_{\T{z}}\) are the dominant co-polarized field components, whereas \(E_{\T{y}}\) is the cross-polarized component. A notable feature is that the sampling points selected for \(E_{\T{x}}\) and \(E_{\T{z}}\) are highly similar: many of them are located at the same or very nearby positions on the observation grid. This indicates that these two dominant field components contain strongly correlated spatial information and exhibit similar modal structures under dual-polarized excitation.

At the same time, the point distributions of all three components are denser in the region close to the LIS and gradually become sparser as the observation distance increases along the \(y\)-direction. This behavior suggests that the field varies more rapidly in the nearby region, which requires denser sampling to capture the dominant modes accurately. Compared with \(E_{\T{x}}\) and \(E_{\T{z}}\), the sampling points selected for the cross-polarized component \(E_{\T{y}}\) are more distinct and less overlapped with the other two sets, although they still follow the same overall trend of being concentrated in the near-LIS region. This difference implies that \(E_{\T{y}}\) provides more complementary information, while \(E_{\T{x}}\) and \(E_{\T{z}}\) mainly reflect similar dominant modal characteristics.

Figure~\ref{fig:DEIM_points_polarization} shows that different field polarizations do not contribute equally to the sampling structure. The co-polarized components \(E_{\T{x}}\) and \(E_{\T{z}}\) share a large portion of their informative sampling locations, revealing a significant degree of redundancy, whereas the \(E_{\T{y}}\) contributes comparatively more independent sampling information. This is consistent with the singular value spectra results, which indicate that different field components are only partially complementary rather than fully independent.

\begin{figure}[t]
  \centering
  \includegraphics[width=0.6\linewidth]{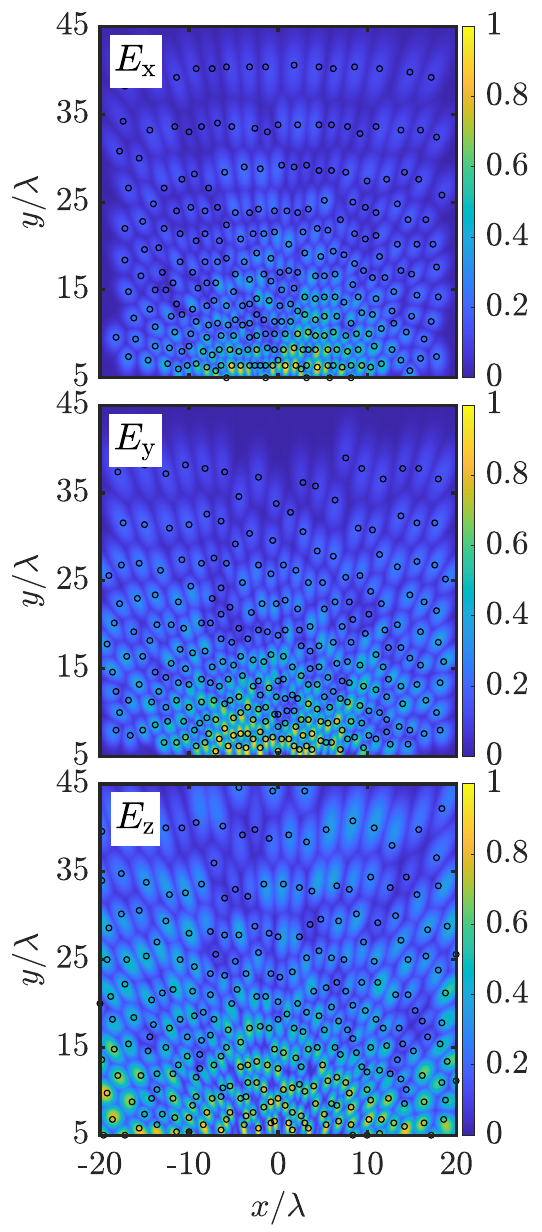}
  \caption{Normalized beamforming intensity maps in the $x$--$y$ plane obtained by focusing on the DEIM-selected sampling points for different electric-field polarizations. The black hollow circles denote the sampling points selected by the DEIM algorithm with beamforming. The simulation parameters are the same as those used in Fig.~\ref{fig:DEIM_points}. The focused beam is designed using the ZF method.}
  \label{fig:3D_surface_sampling_distribution_polarization}
\end{figure}

Figure~\ref{fig:3D_surface_sampling_distribution_polarization} shows the normalized beamforming intensity maps in the \(x\)-\(y\) plane obtained by focusing on the DEIM-selected sampling points for different received electric-field polarizations. A common feature of all three subfigures is that the beamforming intensity is strongest in the near-field region close to the LIS, and then gradually decreases as the observation distance increases along the \(y\)-direction. This is consistent with the DEIM point distributions, where denser sampling is required in the region closer to the LIS in order to capture the faster spatial field variation.

The beam patterns corresponding to different field components are clearly different. For the cross-polarized component \(E_{\T{y}}\), the high-intensity responses are mainly concentrated in the near-LIS region, while several oblique and interference-like streaks extend into the farther observation area. This indicates that \(E_{\T{y}}\) still contains useful spatial information, but its effective focusing region is relatively limited compared with the dominant co-polarized components.

For the co-polarized components \(E_{\T{x}}\) and \(E_{\T{z}}\), the beamforming responses are generally stronger and spread over a broader portion of the observation region. In particular, the intensity map of \(E_{\T{x}}\) exhibits a relatively regular and smooth distribution, with strong responses concentrated around the central near-field region and gradually decaying with distance. The \(E_{\T{z}}\) component shows an even broader spatial spread, together with more pronounced stripe-like and oscillatory structures over the \(x\)-\(y\) plane. This suggests that \(E_{\T{z}}\) captures richer modal variations and remains effective over a wider region. Therefore, although \(E_{\T{x}}\) and \(E_{\T{z}}\) are both dominant co-polarized components, their beamforming characteristics are not identical.

\section{Conclusions}
This paper investigated the spatial DoF of LISs and its connection to sampling design and near-field beamforming. The available DoF is characterized using the mutual shadow area, with representative closed-form results derived for 2D line-source and 3D surface-source configurations. The predicted DoF is validated by numerical singular-value spectra.

Based on this DoF characterization, efficient receiver-plane sampling strategies are developed. Closed-form sampling rules are used for 2D line-source configurations, while DEIM-based nonuniform sampling is applied to 3D surface-source configurations. The selected points are then used for MRT and ZF beamforming evaluation, including polarization-aware field analysis. The results demonstrate that the proposed sampling strategies effectively translate theoretical DoF predictions into practical beamforming performance evaluation.

Finally, polarization effects are analyzed. The polarization-aware results
show that different electric-field components contribute unequally to the
effective DoF. The total-field DoF is approximately twice that of a single
polarization component.

\appendices
\section{Green’s Functions}\label{app:2D_Source}
The scalar Green's functions in $\mathbb{R}^2$ and $\mathbb{R}^3$ are given by~\cite{gustafsson2025shadow} 
\begin{equation}
G_2
=
\frac{\ju}{4}
\T H_0^{(2)}
\left(k\lvert \V r-\V r^\prime \rvert\right), \qquad
G_3
=
\frac{\eu^{-\ju k\lvert \V r-\V r^\prime \rvert}}
{4\pi \lvert \V r-\V r^\prime\rvert},
\label{eq:transmission matrices}
\end{equation}
respectively, where $\T H_0^{(2)}$ is the Hankel function of the second kind and order zero.
The Green's dyadic
\begin{equation}
\M{G_\T J} = \ju k\left( \M I_3 + k^{-2}\nabla\nabla\right) G_3, \qquad
\M{G_\T M} =  \nabla \times (\M I_3 G_3),
\label{eq:dyadic_eq}
\end{equation}
where $\M{G}_\T{J}$ maps the electric current $\V J$ to the electric field,
whereas $\M{G}_\T{M}$ maps the normalized magnetic current
$\eta_0^{-1}\V M$ to the electric field. $k=2\pi/\lambda$ is the free space wavenumber, \(\eta_0\) is the wave impedance in the free space, and $\M I_3$ is the unit dyad.

\bibliographystyle{IEEEtran}
\bibliography{citations}

@article{hu2018beyond1,
  title={Beyond massive {MIMO}: The potential of data transmission with large intelligent surfaces},
  author={Hu, Sha and Rusek, Fredrik and Edfors, Ove},
  journal={IEEE Trans. Signal Process.},
  volume={66},
  number={10},
  pages={2746--2758},
  year={May 2018},
  publisher={IEEE}
}

@article{hu2023design,
  title={Design of near-field beamforming for large intelligent surfaces},
  author={Hu, Sha and Wang, Hao and Ilter, Mehmet C},
  journal={IEEE Trans. Wireless Commun.},
  volume={23},
  number={1},
  pages={762--774},
  year={Jan. 2023},
  publisher={IEEE}
}

@article{selvan2017fraunhofer,
  title={Fraunhofer and {F}resnel distances: Unified derivation for aperture antennas},
  author={Selvan, Krishnasamy T and Janaswamy, Ramakrishna},
  journal={IEEE Antennas Propag.
Mag.},
  volume={59},
  number={4},
  pages={12--15},
  year={Aug. 2017},
  publisher={IEEE}
}

@article{gustafsson2025shadow,
  title={Shadow area and degrees of freedom for free-space communication},
  author={Gustafsson, Mats},
  journal={IEEE J. Sel. Areas Inf. Theory},
  volume  = {6},
  pages   = {325-337},
  year={2025},
  publisher={IEEE}
}

@article{gustafsson2025degrees,
  title={Degrees of freedom for radiating systems},
  author={Gustafsson, Mats},
  journal={IEEE Trans. Antennas Propag.},
  volume={73},
  number={2},
  pages={1028--1038},
  year={Feb. 2025},
  publisher={IEEE}
}

@article{zhang2022beam,
  author={Zhang, Haiyang and Shlezinger, Nir and Guidi, Francesco and Dardari, Davide and Imani, Mohammadreza F. and Eldar, Yonina C.},
  title   = {Beam Focusing for Near-Field Multiuser {MIMO} Communications},
  journal = {IEEE Trans. Wireless Commun.},
  volume  = {21},
  number  = {9},
  pages   = {7476--7490},
  year    = {Sep. 2022},
  doi     = {10.1109/TWC.2022.3158894}
}

@article{li2024multiuser,
   author={Li, Xinrui and Dong, Zhenjun and Zeng, Yong and Jin, Shi and Zhang, Rui},
  title   = {Multi-User Modular {XL-MIMO} Communications: Near-Field Beam Focusing Pattern and User Grouping},
  journal = {IEEE Trans. Wireless Commun.},
  volume  = {23},
  number  = {10},
  pages   = {13766--13781},
  year    = {Oct. 2024},
  doi     = {10.1109/TWC.2024.3404659}
}

@article{chen2024beamspace,
  author  = {Chen, Kangjian and Qi, Chenhao and Huang, Jingjia and Dobre, Octavia A. and Li, Geoffrey Ye},
  title   = {Near-Field Communications for Extremely Large-Scale {MIMO}: A Beamspace Perspective},
  journal = {IEEE Commun. Mag.},
  volume={63},
  number={5},
  pages={166--172},
  year={May 2025},
  publisher={IEEE}
}

@article{cui2023nearfieldrainbow,
  author  = {Cui, M. and Dai, L. and Zhang, R.},
  title   = {Near-Field Rainbow: Wideband Beam Training for {XL-MIMO}},
  journal = {IEEE Trans. Wireless Commun.},
  volume  = {22},
  number  = {6},
  pages   = {3899--3912},
  year    = {Jun. 2023},
  doi     = {10.1109/TWC.2022.3226814}
}

@book{weyl1911asymptotische,
  author    = {Arendt, W. and Nittka, R. and Peter, W. and Steiner, F.},
  title     = {Weyl's Law: Spectral Properties of the Laplacian in Mathematics and Physics},
  publisher = {John Wiley \& Sons, Ltd},
  year      = {2009},
  chapter   = {1},
  pages     = {1--71}
}

@article{miller2000communicating,
  author  = {Miller, David AB},
  title   = {Waves, modes, communications, and optics: a tutorial},
  journal = {Advances in Optics and Photonics},
  volume  = {11},
  number  = {3},
  pages   = {679--825},
  year    = {2019},
  doi     = {10.1364/AO.39.001681}
}

@article{kuang2025bounds,
  title={Bounds on the coupling strengths of communication channels and their information capacities},
  author={Kuang, Zeyu and Miller, David AB and Miller, Owen D},
  journal={IEEE Trans. Antennas Propag.},
  volume={73},
  number={6},
  pages={3959--3974},
  year={Jan. 2025},
  publisher={IEEE}
}

@article{migliore2006role,
  author  = {Migliore, M. D.},
  title   = {On the Role of the Number of Degrees of Freedom of the Field in {MIMO} Channels},
  journal = {IEEE Trans. Antennas Propag.},
  volume  = {54},
  number  = {2},
  pages   = {620--628},
  year    = {Feb. 2006},
  doi     = {10.1109/TAP.2005.863398}
}

@article{jensen2008capacity,
  author  = {Jensen, M. A. and Wallace, J. W.},
  title   = {Capacity of the Continuous-Space Electromagnetic Channel},
  journal = {IEEE Trans. Antennas Propag.},
  volume  = {56},
  number  = {2},
  pages   = {524--531},
  year    = {Feb. 2008},
  doi     = {10.1109/TAP.2007.915421}
}

@article{bucci1989degrees,
  author  = {Bucci, O. M. and Franceschetti, G.},
  title   = {On the Degrees of Freedom of Scattered Fields},
  journal = {IEEE Trans. Antennas Propag.},
  volume  = {37},
  number  = {7},
  pages   = {918--926},
  year    = {Jul. 1989},
  doi     = {10.1109/8.29386}
}

@article{maisto2021near,
  title={Near-field transverse resolution in planar source reconstructions},
  author={Maisto, Maria Antonia and Pierri, Rocco and Solimene, Raffaele},
  journal={IEEE Trans. Antennas Propag.},
  volume={69},
  number={8},
  pages={4836--4845},
  year={Aug. 2021},
  publisher={IEEE}
}

@article{puggelli2025maximizing,
  title={Maximizing independent channels and efficiency in {BTS} array antennas via {EM} degrees of freedom},
  author={Puggelli, Federico and Biscontini, Bruno and Martini, Enrica and Maci, Stefano},
  journal={IEEE Trans. Antennas Propag.},
  year={Jun. 2025},
  volume={73},
  number={6},
  pages={3444-3458},
  publisher={IEEE}
}

@ARTICLE{Bucci2025,
  author={Bucci, Ovidio Mario and Migliore, Marco Donald},
  journal={IEEE Antennas Propag. Mag.}, 
  title={Degrees of Freedom and Sampling Representation of Electromagnetic Fields: Concepts and applications.}, 
  year={Jun. 2025},
  volume={67},
  number={3},
  pages={10-22},
  doi={10.1109/MAP.2024.3513216}}

@article{bucci1998representation,
  author  = {Bucci, O. M. and Gennarelli, C. and Savarese, C.},
  title   = {Representation of Electromagnetic Fields over Arbitrary Surfaces by a Finite and Nonredundant Number of Samples},
  journal = {IEEE Trans. Antennas Propag.},
  volume  = {46},
  number  = {3},
  pages   = {351--359},
  year    = {Mar. 1998},
  doi     = {10.1109/8.662654}
}

@article{franceschetti2009mimo,
  author  = {Franceschetti, M. and Migliore, M. D. and Minero, P.},
  title   = {The Capacity of Wireless Networks: Information-Theoretic and Physical Limits},
  journal = {IEEE Trans. Inf.
Theory},
  volume  = {55},
  number  = {8},
  pages   = {3413--3424},
  year    = {Aug. 2009},
  doi     = {10.1109/TIT.2009.2023705}
}

@book{franceschetti2017wave,
  author    = {Franceschetti, M.},
  title     = {Wave Theory of Information},
  publisher = {Cambridge Univ. Press},
  address   = {Cambridge, U.K.},
  year      = {2017}
}

@article{bjornson2024towards,
  title={Towards {6G MIMO}: Massive spatial multiplexing, dense arrays, and interplay between electromagnetics and processing},
  author={Bj{\"o}rnson et al., Emil},
  journal={arXiv preprint arXiv:2401.02844},
  year={2024}
}

@article{dardari2020communicating,
  title={Communicating with large intelligent surfaces: Fundamental limits and models},
  author={Dardari, Davide},
  journal={IEEE J. Sel. Areas Commun.},
  volume={38},
  number={11},
  pages={2526--2537},
  year={Nov. 2020},
  publisher={IEEE}
}

@book{horn2012matrix,
  title={Matrix analysis},
  author={Horn, Roger A and Johnson, Charles R},
  year={2012},
  publisher={Cambridge Univ. Press}
}

@article{brick2026interpreting,
  title={Interpreting Moment Matrix Blocks Spectra using Mutual Shadow Area},
  author={Brick, Yaniv and Andriulli, Francesco P and Gustafsson, Mats},
  journal={IEEE Trans. Antennas Propag.  (in press), arXiv:2601.17965},
  year={2026}
}

@article{christensen2008weighted,
  title={Weighted Sum-Rate Maximization Using Weighted {MMSE} for {MIMO-BC} Beamforming Design},
  author={Christensen, S{\o}ren Skovgaard and Agarwal, Rajiv and De Carvalho, Elisabeth and Cioffi, John M},
  journal={IEEE Trans. Wireless Commun.},
  volume={7},
  number={12},
  pages={4792--4799},
  year={Dec. 2008},
  publisher={IEEE}
}

@article{decarli2021communication,
  author  = {Decarli, N. and Dardari, D.},
  title   = {Communication Modes with Large Intelligent Surfaces in the Near Field},
  journal = {IEEE Access},
  volume  = {9},
  pages   = {165648--165666},
  year    = {Dec. 2021},
  doi     = {10.1109/ACCESS.2021.3133707}
}

@article{yuan2024breaking,
  author={Yuan et al., Shuai S. A.},
  title   = {Breaking the Degrees-of-Freedom Limit of Holographic {MIMO} Communications: A {3-D} Antenna Array Topology},
  journal = {IEEE Trans.
Veh. Technol.},
  volume  = {73},
  number  = {8},
  pages   = {11276--11288},
  year    = {Aug. 2024},
}

@article{maisto2021efficient,
  author  = {Maisto, M. A. and Leone, G. and Brancaccio, A. and Solimene, R.},
  title   = {Efficient Planar Near-Field Measurements for Radiation Pattern Evaluation by a Warping Strategy},
  journal = {IEEE Access},
  volume  = {9},
  pages   = {62255--62265},
  year    = {Apr. 2021},
  doi     = {10.1109/ACCESS.2021.3074786}
}

@article{howell2011radiative,
  title={Radiative transfer configuration factor catalog: A listing of relations for common geometries},
  author={Howell, John R and Meng{\"u}{\c{c}}, M Pinar},
  journal={J. Quant. Spectrosc. Radiat. Transf.},
  volume={112},
  number={5},
  pages={910--912},
  year={2011},
  publisher={Elsevier}
}

@article{rao2001performance,
  title={Performance of maximal ratio transmission with two receive antennas},
  author={Rao, Bhaskar D and Yan, Ming},
  journal={IEEE Trans. Commun.},
  volume={51},
  number={6},
  pages={894--895},
  year={Jun. 2003},
  publisher={IEEE}
}

@article{shen2020wireless,
  title={Wireless power transfer with distributed antennas: System design, prototype, and experiments},
  author={Shen, Shanpu and Kim, Junghoon and Song, Chaoyun and Clerckx, Bruno},
  journal={IEEE Trans. Ind. Electron.},
  volume={68},
  number={11},
  pages={10868--10878},
  year={Nov.2020},
  publisher={IEEE}
}

@article{spencer2004zero,
  title={Zero-forcing methods for downlink spatial multiplexing in multiuser {MIMO} channels},
  author={Spencer, Quentin H and Swindlehurst, A Lee and Haardt, Martin},
  journal={IEEE Trans. Signal Process.},
  volume={52},
  number={2},
  pages={461--471},
  year={Feb. 2004},
  publisher={IEEE}
}

@article{hochman2014reduced,
  title={Reduced-order models for electromagnetic scattering problems},
  author={Hochman, Amit and Villena, Jorge Fernández and Polimeridis, Athanasios G. and Silveira, Luís Miguel and White, Jacob K. and Daniel, Luca},
  journal={IEEE Trans. Antennas Propag.},
  volume={62},
  number={6},
  pages={3150--3162},
  year={Jun. 2014},
  publisher={IEEE}
}

@article{poon2005degrees,
  title={Degrees of freedom in multiple-antenna channels: A signal space approach},
  author={Poon, Ada SY and Brodersen, Robert W and Tse, David NC},
  journal={IEEE Trans. Inf.
Theory},
  volume={51},
  number={2},
  pages={523--536},
  year={Feb. 2005},
  publisher={IEEE}
}

\vfill

\end{document}